%% file: main.tex
\documentclass{vldb}
\usepackage{graphicx}
\usepackage{mdwlist}
\usepackage{tabularx}
\usepackage{balance}  
\usepackage{verbatim}
\usepackage{graphicx}
\usepackage[export]{adjustbox}
\usepackage{listings}
\usepackage{latexsym}
\usepackage{url}
\usepackage[usenames,dvipsnames]{color}
\usepackage{multirow} 
\usepackage{subfigure} 
\usepackage{listings}
\usepackage{appendix}
\newcommand{\timFutureToDo}[1]{}

\begin{document}
\setlength\abovedisplayskip{0pt plus 2pt minus 0pt}
\setlength\belowdisplayskip{0pt plus 2pt minus 0pt}

\makeatletter
\g@addto@macro \normalsize {%
 \setlength\abovedisplayskip{0pt plus 2pt minus 0pt}%
 \setlength\belowdisplayskip{0pt plus 2pt minus 0pt}%
 \setlength\abovedisplayshortskip{0pt plus 2pt minus 0pt}%
 \setlength\belowdisplayshortskip{0pt plus 2pt minus 0pt}%
}
\makeatother

\makeatletter
\g@addto@macro \small {%
 \setlength\abovedisplayskip{0pt plus 2pt minus 0pt}%
 \setlength\belowdisplayskip{2pt plus 2pt minus 0pt}%
  \setlength\abovedisplayshortskip{0pt plus 2pt minus 0pt}%
 \setlength\belowdisplayshortskip{2pt plus 2pt minus 0pt}%
}
\makeatother

\newenvironment{packed_item}{
\begin{list}{$\bullet$}{
  \setlength{\itemsep}{-2pt}
  \setlength{\parskip}{1pt}
  \setlength{\labelwidth}{15 pt}
  \setlength{\leftmargin}{10pt}
  \setlength{\itemindent}{0pt}}
}{\end{list}}

\title{The End of Slow Networks:\\ It's Time for a Redesign [Vision]}

\author{
\begin{tabular}{ccccc}\\
Carsten Binnig  & Andrew Crotty & Alex Galakatos & Tim Kraska & Erfan Zamanian   \\
\end{tabular}\\
\begin{tabular}{c}\\
\eaddfnt{\normalsize Brown University,} 
\eaddfnt{\normalsize firstname\_lastname@brown.edu}
\end{tabular}}


\maketitle
\vspace*{-70pt}
\begin{abstract}
The next generation of high-performance RDMA-capable networks requires a fundamental rethinking of the design of modern distributed in-memory DBMSs.
These systems are commonly designed under the assumption that the network is the bottleneck and thus must be avoided as much as possible. This assumption no longer holds true. 
With InfiniBand FDR 4x, the bandwidth available to transfer data across the network is in the same ballpark as the bandwidth of one memory channel, and the bandwidth increases even more with the most recent EDR standard.
Moreover, with increasing advances in RDMA, transfer latencies improve similarly fast.
In this paper, we first argue that the ``old'' distributed database design is not capable of taking full advantage of fast networks and suggest a new architecture. 
Second, we discuss initial results of a prototype implementation of this architecture for OLTP and OLAP, and show remarkable performance improvements over existing designs.
\end{abstract}

\vspace{-2ex}
\input{intro.tex}
\vspace{-2ex}
\input{background.tex}
\vspace{-2ex}
\input{architecture.tex}
\vspace{-2ex}
\input{oltp.tex}
\vspace{-2ex}
\input{olap.tex}
\vspace{-2ex}
\input{related.tex}
\vspace{-2ex}
\input{concl.tex}
\vspace{-2ex}

\begin{scriptsize}
\bibliographystyle{abbrv}
\bibliography{bib}
\end{scriptsize}


\end{document}

%% file: intro.tex
\section{Introduction}
\label{sec:intro}
We argue that the current trend towards high-performance \textit{Remote Direct Memory Access} (RDMA) capable networks, such as InfiniBand FDR/EDR, will require a complete redesign of modern distributed in-memory DBMSs, which are built on the assumption that the network is the main bottleneck~\cite{BabuSurvey}.
Consequently, these systems aim to avoid communication between machines, using techniques such as locality-aware partitioning schemes~\cite{Sword:EDBT:2013,HStore:SIGMOD:2012,Schism:VLDB:2010, LocalityAware}, semi-reductions for joins~\cite{bloom}, and complicated preprocessing steps~\cite{trackj,roediger}.
Yet, with the nascent modern network technologies, the assumption that the network is the bottleneck no longer holds. 

\begin{figure}[t]
\begin{center}
\vspace*{-26pt}
\subfigure[Specification]{
   \hspace{-9ex}
   \includegraphics[trim = 0mm 0mm 0mm 0mm, width=.26\textwidth]{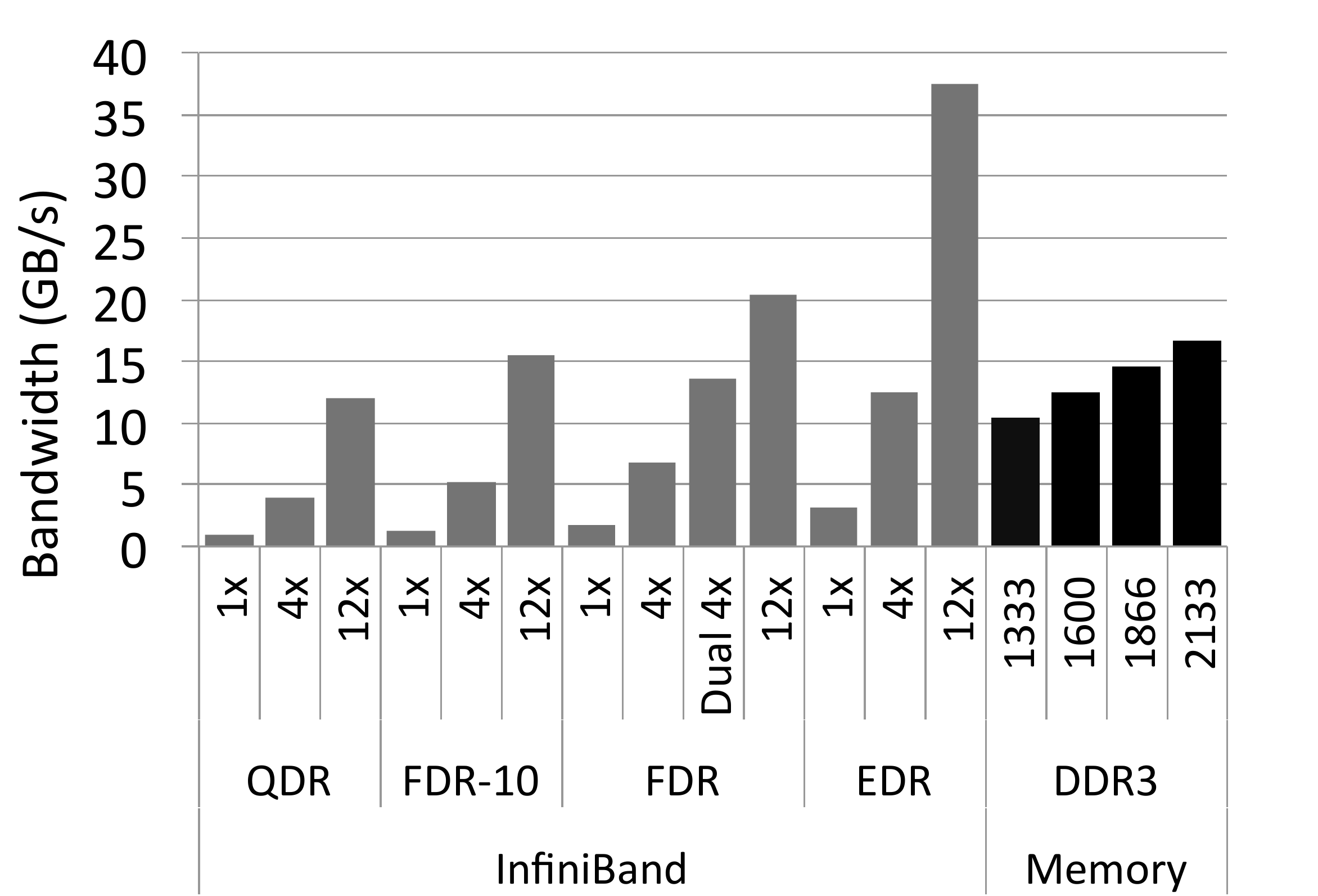}
   \label{fig:bandwidth_mem_ib:spec}
 }
 \subfigure[Experiment]{
   \hspace{-4ex}
   \includegraphics[trim = 0mm -5mm 33mm 30mm, width=.16\textwidth]{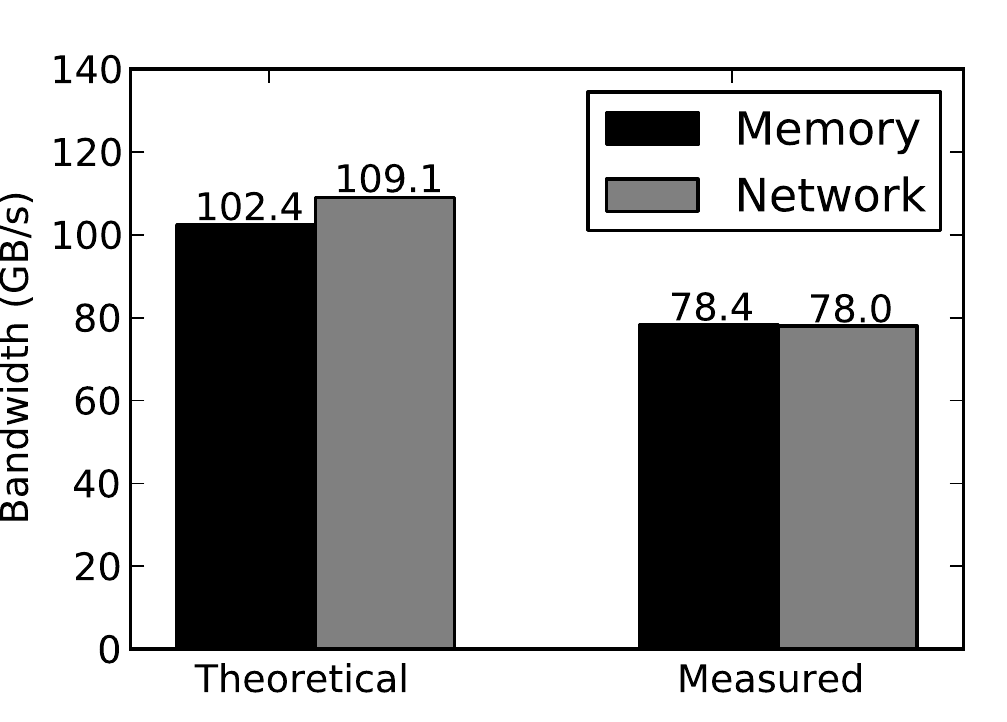}
   \label{fig:bandwidth_mem_ib:setup}
}
\vspace*{-12pt}
\caption{Memory vs Network Bandwidth: (a) specification, (b) for a Dual-socket Xeon E5v2 server with DD3-1600 and two FDR 4x NICs per socket} 
\vspace*{-28pt}
\end{center}
\label{fig:bandwidth_mem_ib}
\end{figure}

Even today, with InfiniBand FDR 4$\times$~\cite{IBSpec}, the bandwidth available to transfer data across the network is in the same ballpark as the bandwidth of one memory channel.
DDR3 memory bandwidth currently ranges from 6.25 GB/s (DDR3-800) to 16.6 GB/s (DDR3-2133) \cite{ddr3-standard} per channel, whereas InfiniBand has a specified bandwidth of 1.7 GB/s (FDR 1$\times$) to 37.5GB/s (EDR 12$\times$) \cite{IBSpec} per NIC port (see Figure~\ref{fig:bandwidth_mem_ib:spec}).
Moreover, future InfiniBand standards (HDR as well as NDR) promise a bandwidth that exceeds the bandwidth of the local memory bus by far.

However, modern systems typically support 4 memory channels per socket.
For example, a machine with DDR3-1600 memory has 12.8GB/s per channel, with a total aggregate memory bandwidth of 51.2GB/s, and 4 dual-port FDR $4\times$ NICs provide roughly the same bandwidth.\footnote{We do \emph{not} assume that the PCIe bus becomes a bottleneck, as current dual socket Xeon e5 boards typically have 40 Gen3 lanes per socket, achieving 39.4 GB/s total bandwidth.}
Even more surprisingly, the CPU-memory bandwidth is half-duplex, while InfiniBand and PCIe are full-duplex, such that \textit{only 2 NICs} could saturate the memory bandwidth of a read/write workload.
Figure~\ref{fig:bandwidth_mem_ib:setup} shows the theoretical (left) and measured (right) total memory and network throughput for a dual-socket machine with DDR3-1600 memory and two FDR 4$\times$ NICs per socket (4 in total).
This microbenchmark shows that the network transfer is indeed limited by the total available memory bandwidth, not the network bandwidth (see also Section~\ref{sec:background} for more microbenchmarks).
While these measures were done for InfiniBand, we expect that Ethernet networks will become similarly advanced \cite{RDMAEthernet,RoCEIB,RoCEStory}.

Another important factor is that with major advances in RDMA, the network latency also improves quickly.
Our recent experiments with InfiniBand FDR 4$\times$ showed that the system requires  $\approx 1\mu$s to transfer $1$KB of data using RDMA, compared to $\approx 0.08\mu$s for the CPU to read the same amount of data from memory. 
With only $256$KB, there is virtually no difference between the access time since the bandwidth starts to dominate the transfer time.
Yet, we do not argue that the network latency will become as fast as the memory latency. 
Instead, cache- and memory-locality will play an even more important role for small data requests (e.g., a hash-table look-up) as the system performance is no longer dominated by the network transfer time. 

At the same time, particularly for smaller deployments, InfiniBand is becoming more affordable. 
For example, a small cluster with $8$ servers, 2$\times$ Xeon E5v2 CPUs per machine, $2$~TB of DDR3-1600 memory, and one 2-port InfiniBand FDR 4$\times$ NIC per machine costs under \$80K, with roughly \$20K for the switch and NICs. 
In this configuration, the bandwidth for sending data across the network is close to the bandwidth of one memory channel (13.6 GB/s for network vs. 12.8 GB/s for memory).
Furthermore, memory prices continue to drop, making it feasible to keep even large data sets entirely in memory with just a few machines~\cite{memoryprices}, removing the disk as a bottleneck and created a more balanced system.

However, it is {\em wrong to assume that the fast network changes the cluster to a NUMA architecture} because: (1) the RDMA-based memory access patterns are very different from a local memory access in a NUMA architecture; (2) the latency between machines is still higher to access a single (random) byte than with today's NUMA systems; and (3) hardware-embedded coherence mechanisms ensure data consistency in a NUMA architecture, which is not supported with RDMA.
Clusters with RDMA-capable networks are most similar to a hybrid shared-memory and message-passing system: it is neither a shared-memory system (several address spaces exist) nor a pure message-passing system (data can be directly accessed via RDMA). 

Consequently, we believe there is a need to critically rethink the entire distributed DBMS architecture to take full advantage of the next generation of network technology. 
For example, given the fast network, it is no longer obvious that avoiding distributed transactions is always beneficial.
Similarly, distributed storage managers and distributed execution algorithms (e.g., joins) should no longer be designed to avoid communication at all costs \cite{trackj}, but instead should consider the multi-core architecture and caching effects more carefully even in the distributed environment.
While this is not the first attempt to leverage RDMA for databases \cite{OLAPRamCloud,hyperrdma15,shared-database:sigmod2015}, existing work does not fully recognize that next generation networks create an architectural 4 point. 

This paper makes the following contributions:

\vspace{-2ex}
\begin{itemize*}
\item We present microbenchmarks to assess performance characteristics of one of the latest InfiniBand standards, FDR 4x (Section~\ref{sec:background}).
\item We present alternative architectures for a distributed in-memory DBMS over fast networks and introduce a novel Network-Attached Memory (NAM) architecture (Section~\ref{sec:architecture}).
\item We show why the common wisdom that says ``2-phase-commit does not scale'' no longer holds true for RDMA-enabled networks and outline how OLTP workloads can take advantage of the network by using the NAM architecture. (Section~\ref{sec:oltp})
\item We analyze the performance of distributed OLAP operations (joins and aggregations) and propose new algorithms for the NAM architecture (Section~\ref{sec:olap}). 
\end{itemize*}
\vspace*{-2ex}



%% file: background.tex
\section{Background}
\label{sec:background}
Before making a detailed case why distributed DBMS architectures need to fundamentally change to take advantage of the next generation of network technology, we provide some background information and micro-benchmarks that showcase the characteristics of InfiniBand and RDMA.

\vspace*{-3pt}
\subsection{InfiniBand and RDMA}
\label{sec:background:ib}
In the past, InfiniBand was a very expensive, high bandwidth, low latency network commonly found in large high-performance computing environments.
However, InfiniBand has recently become cost-competitive with Ethernet and thus a viable alternative for enterprise clusters.

\noindent\textbf{Communication Stacks:}
InfiniBand offers two network communication stacks: IP over InfiniBand (IPoIB) and Remote Direct Memory Access (RDMA).
IPoIB implements a classic TCP/IP stack over InfiniBand, allowing existing socket-based applications to run without modification.
As with Ethernet-based networks, data is copied by the application into OS buffers, and the kernel processes the buffers by transmitting packets over the network. 
While providing an easy migration path from Ethernet to InfiniBand, our experiments show that IPoIB cannot fully leverage the network. 
On the other hand, RDMA provides a \textit{verbs} API, which enable data transfers using the processing capabilities of an RDMA NIC (RNIC).
With verbs, most of the processing is executed by the RNIC without OS involvement, which is essential for achieving low latencies.

RDMA provides two verb communication models: one-sided and two-sided. 
One-sided RDMA verbs (write, read, and atomic operations) are executed without involving the CPU of the remote machine.
RDMA WRITE and READ operations allow a machine to write (read) data into (from) the remote memory of another machine.
Atomic operations ({\em fetch-and-add, compare-and-swap}) allow remote memory to be modified atomically.
Two-sided verbs (SEND and RECEIVE) enable applications to implement an RPC-based communication pattern that resembles the socket API.
Unlike the first category, two-sided operations involve the CPU of the remote machine as well.

\noindent \textbf{RDMA Details:}
RDMA connections are implemented using queue pairs (i.e., send/receive queues).
The application creates the queue pairs on the client and the server and the RNICs handle the state of the queue pairs.
To communicate, a client creates a Work Queue Element (WQE) by specifying the verb and parameters (e.g., a remote memory location).
The client puts the WQE into a send queue and informs the local RNIC via Programmed IO (PIO) to process the WQE.
WQEs can be sent either signaled or unsignaled. 
Signaled means that the local RNIC pushes a completion event into a client's {\em completion queue} (CQ) via a DMA write once the WQE has been processed by the remote side. 
For one-sided verbs, the WQEs are handled by the remote RNIC without interrupting the remote CPU using a DMA operation on the remote side (called server). 
However, as a caveat when using one-sided operations, a memory region must be registered to the local and remote RNIC to be accessible by DMA operations (i.e., the RNIC stores the virtual to physical page mappings of the registered region).
For two-sided verbs, the server does not need to register a memory region, but it must put a RECEIVE request into its receive queue to handle a SEND request from the client.


Since queue pairs process their WQEs in FIFO order, a typical pattern to reduce the overhead on the client side and to hide latency is to use selective signaling.
That is, for send/receive queues of length $n$, the client can send $n-1$ WQEs unsignaled and the $n$-th WQE signaled.
Once the completion event (i.e., the acknowledgment message of the server) for the $n$-th WQE arrives, the client implicitly knows that the previous $n-1$ WQEs have also been successfully processed.
That way, computation and communication on the client can be efficiently overlapped without expensive synchronization mechanisms.

Another interesting aspect is how RDMA operations of an RNIC interfere with operations of the CPU if data is concurrently accessed:
(1) Recent Intel CPUs (Intel Sandy-Bridge and later) provide a feature called Data Direct I/O (DDIO) \cite{intel:ddio}.
With DDIO the DMA executed by the RNIC to read (write) data from (to) remote memory places the data directly in the CPU L3 cache if the memory address is resident in the cache to guarantee coherence.
(2) On other systems the cache is flushed/invalidated by the DMA operation to guarantee coherence.
(3) Finally, non-coherent systems leave the coherency problem to the software. 
These effects must be considered when designing distributed RDMA-based algorithms.
Also note that this only concerns coherence between the cache and memory, not the coherence between data once copied, which is always left to the software. 



\begin{figure}[t]
\begin{center}
\subfigure[Throughput]{
   \hspace{-7ex}
   \includegraphics[width=.25\textwidth]{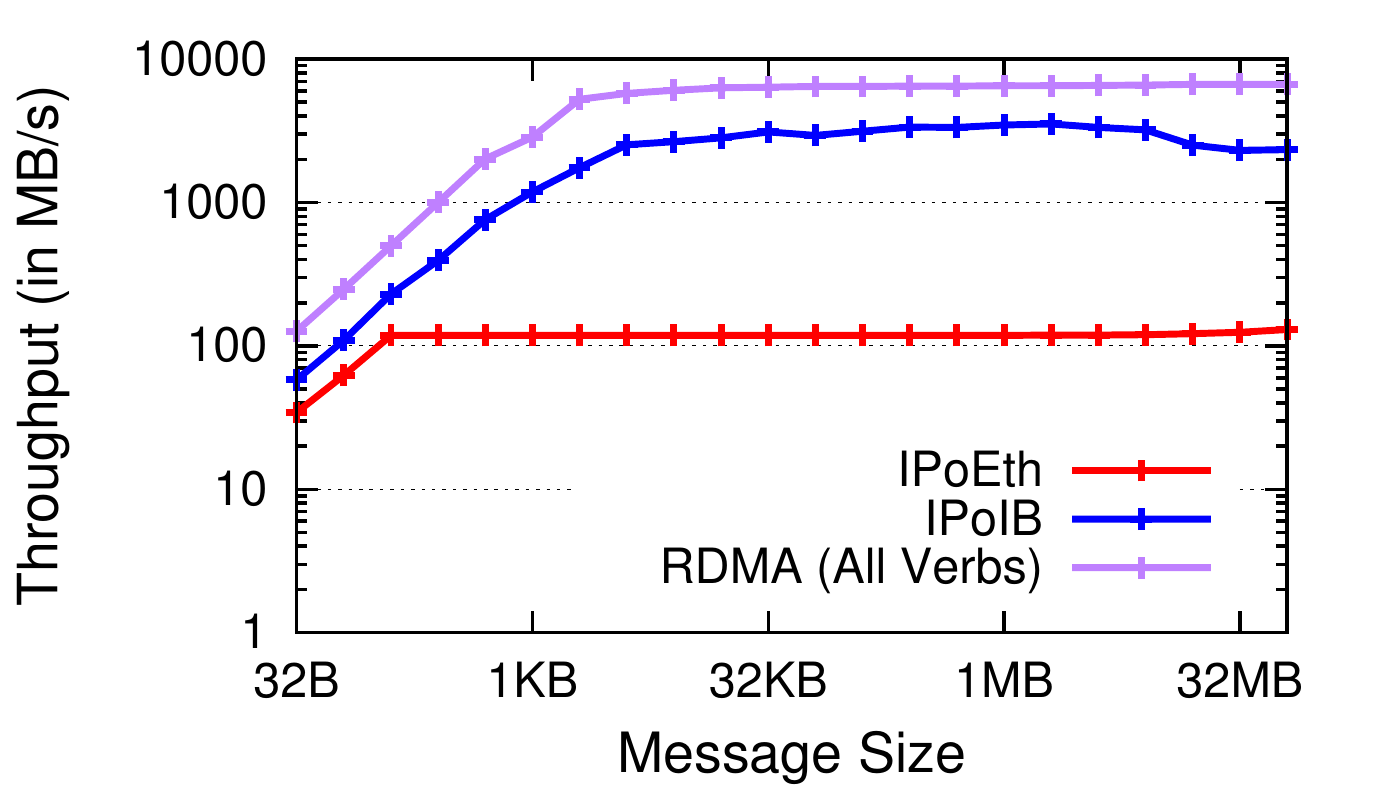}
   \label{fig:exp2bw}
 }
 \subfigure[Latency]{
   \hspace{-3ex}
   \includegraphics[width=.25\textwidth]{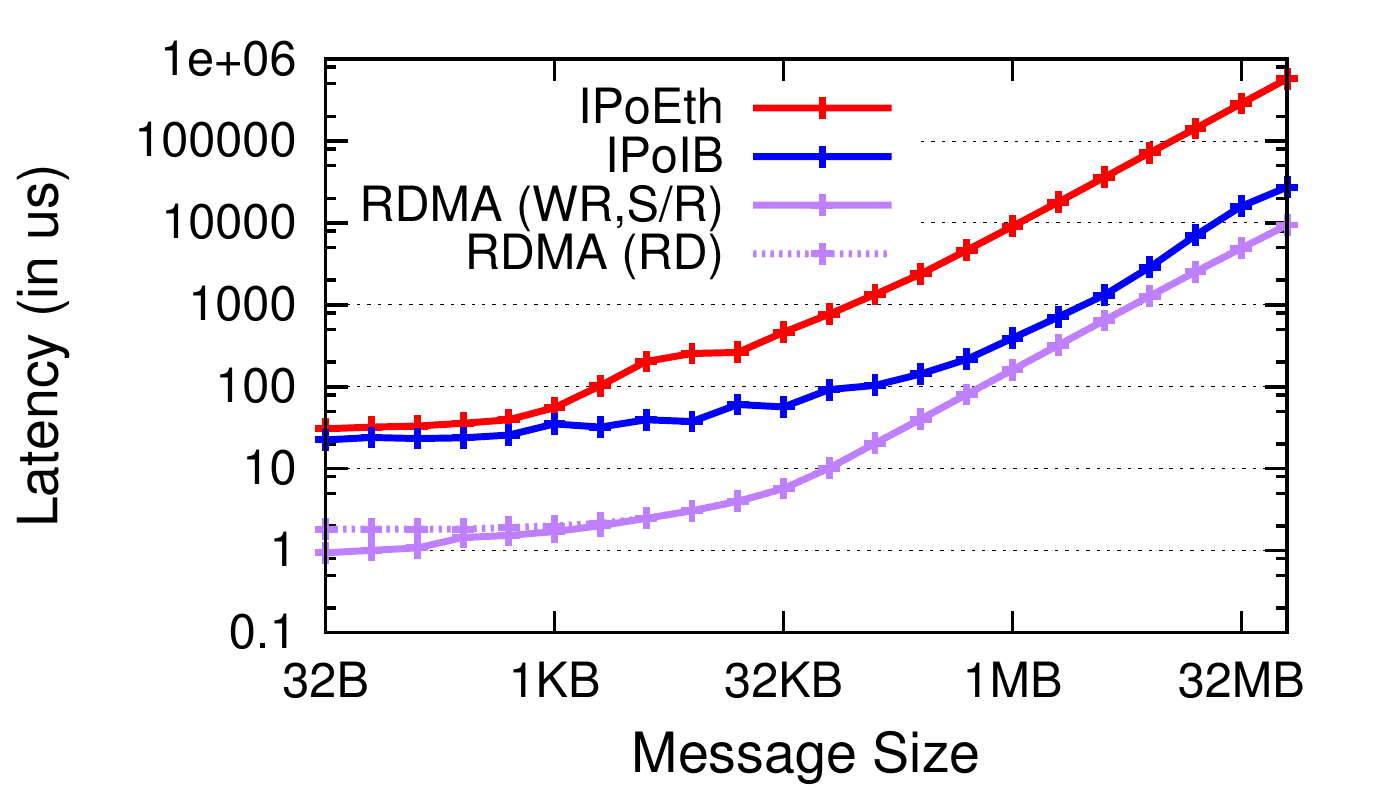}
   \label{fig:exp2lat}
}
\vspace*{-12pt}
\caption{Network Throughput and Latency}
\label{fig:exp:exp2a}
\vspace*{-25pt}
\end{center}
\end{figure}

\vspace*{-3pt}
\subsection{Micro-Benchmarks}
\label{sec:background:micro}
This section presents microbenchmarks that compare the throughput and latency of: (1) a TCP/IP stack over 1Gbps Ethernet (IPoEth), (2) IPoIB, and (3) RDMA.
These results inform the suggestions we make for the redesign of distributed DBMSs on InfiniBand.

\noindent \textbf{Experimental Setup:} 
In our micro-benchmarks we used two machines, each with an Intel Xeon E5-2660 v2 processor and $256$GB RAM.
Both machines were equipped with a Mellanox Connect IB FDR 4x dualport RNIC.
Each port of the RNIC has a bandwidth of $54.54$Gbps ($6.8$GB/s) and is full-duplex.
Additionally, each machine had a 1Gbps Ethernet NIC (with one port) connected to the same Ethernet switch.
Each machine ran Ubuntu Server 14.04 and uses the OFED 2.3.1 driver for the RNIC.

In our experiments, we used one port on the RNIC to better compare the InfiniBand results to the Ethernet results. 
In order to isolate low-level network properties, these microbenchmarks were executed in single-threaded mode.

\noindent\textbf{Throughput and Latency (Figure \ref{fig:exp:exp2a}):} 
For this experiment, we varied the message size from $32$B up to $32$MB to simulate the characteristics of different workloads (OLTP and OLAP) and measured the throughput and latency for IPoEth, IPoIB, and RDMA send/receive and write/read. 
In addition, we also measured the RDMA atomic operations, but since they only support a maximal message size of $8$B and show the same latency and throughput as $8$B READs, we omitted the results from the figure.

\begin{figure}[t]
\begin{center}
\subfigure[Client]{
   \hspace{-7ex}
   \includegraphics[width=.25\textwidth]{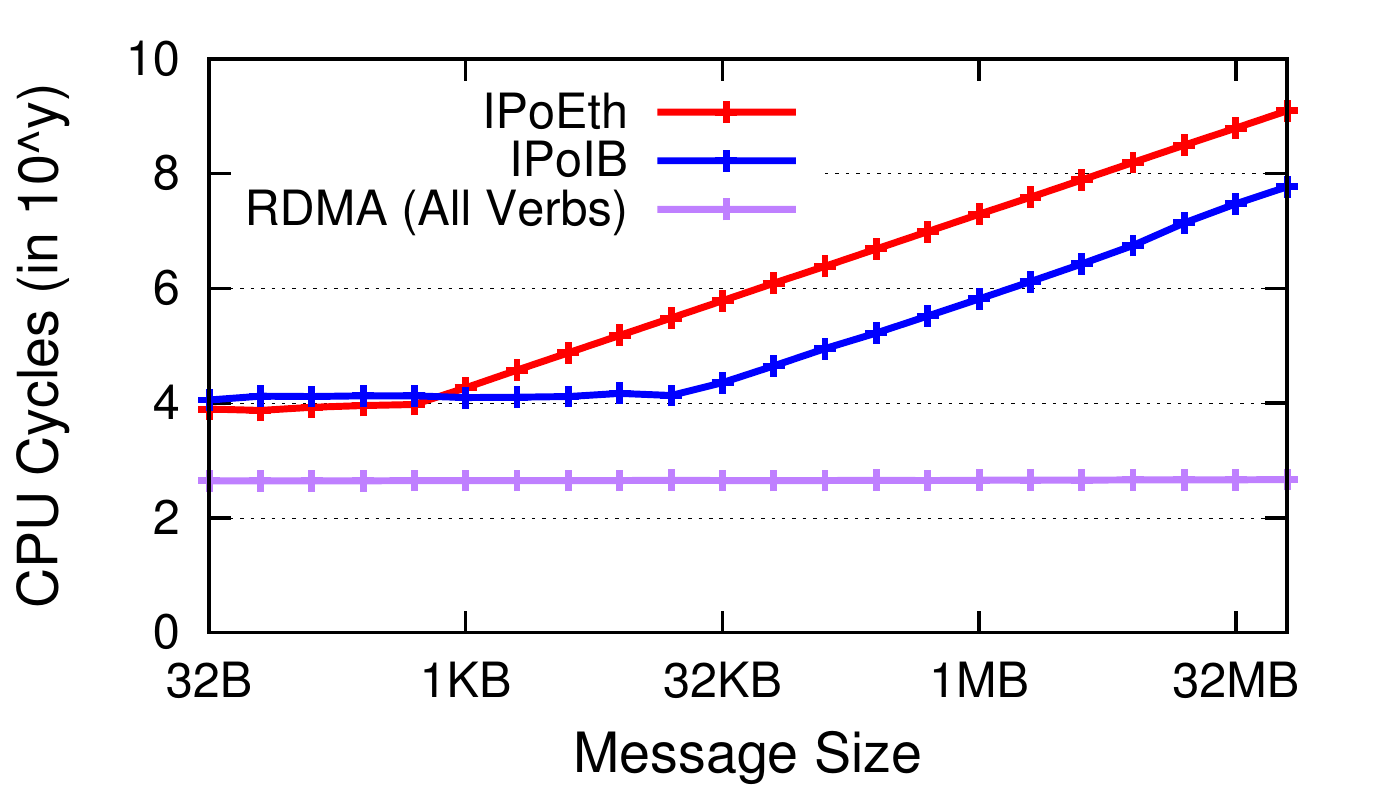}
   \label{fig:exp2oc}
 }
 \subfigure[Server]{
   \hspace{-3ex}
   \includegraphics[width=.25\textwidth]{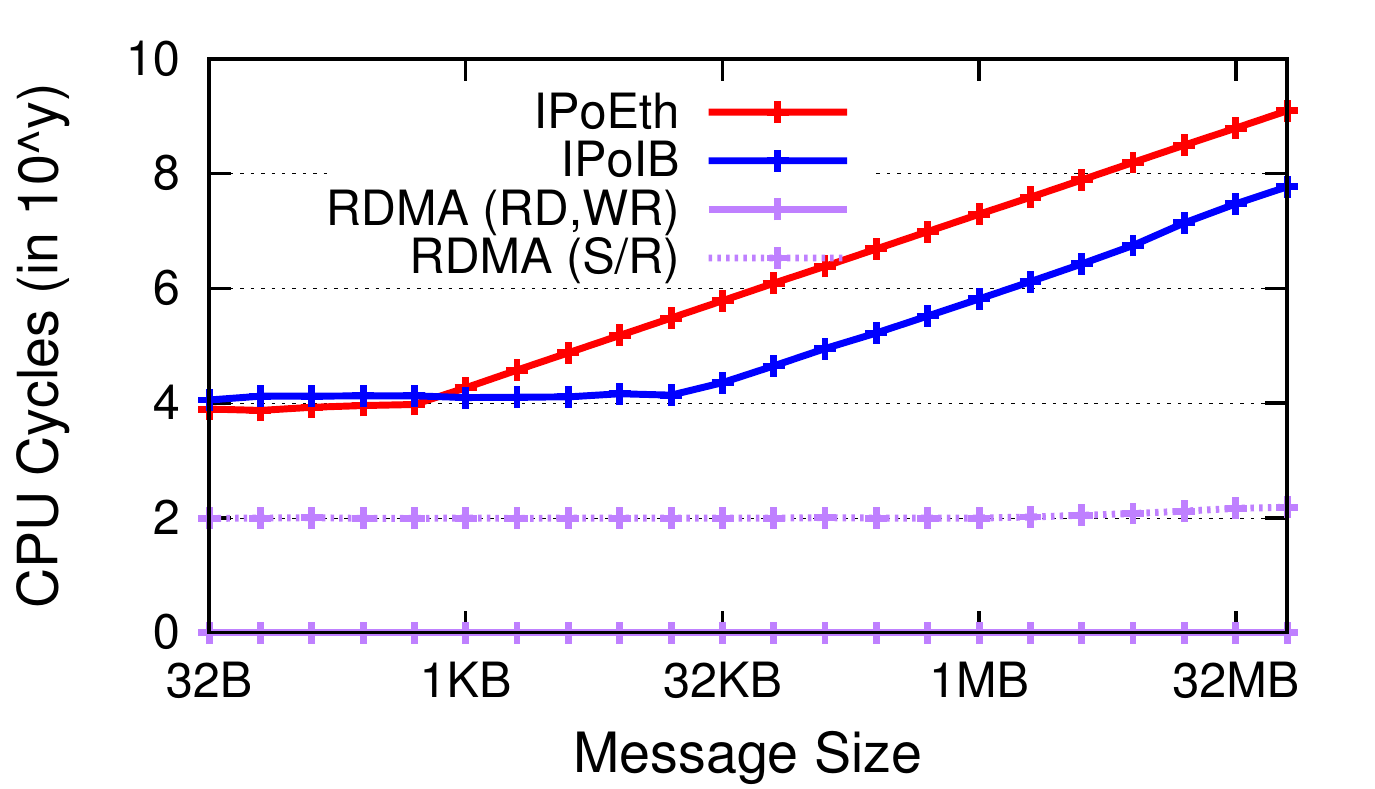}
   \label{fig:exp2os}
}
\vspace*{-12pt}
\caption{CPU Overhead for Network Operations}
\label{fig:exp:exp2b}
\vspace*{-25pt}
\end{center}
\end{figure}

While all RDMA verbs saturate the InfiniBand network bandwidth of approximately $6.8$GB/s for message sizes greater than $2$KB, IPoIB only achieves a maximum throughput of $3.5$GB/s, despite using the same InfiniBand hardware as RDMA.
Moreover, the latency of a message (i.e., 1/2 RTT) over IPoIB is also higher than for RDMA.
In fact, for small message sizes, the latency of IPoIB is much closer to the latency of the 1Gbps Ethernet network (IPoEth).
For example, for a message size of $8$B, the latency is $20\mu$s for IPoIB and $30\mu$s for IPoEth while an RDMA WRITE only takes $1\mu$s.
This is because the TCP/IP stack for IPoIB has a very high CPU overhead per message for small messages (as we will show later in Figure \ref{fig:exp:exp2b}).
For larger message sizes ($\ge 1$MB), the latency of IPoIB is closer to RDMA; however, it is still a factor of $2.5\times$ higher than for RDMA.
For example, a $1$MB message has a latency of $393\mu$s on IPoIB while it has only $161\mu$s for RDMA.

An interesting result is that an RDMA WRITE and a SEND take only $1\mu$s for message sizes less than $256$B while a RDMA READ needs $2\mu$s.
This is because for WRITEs and SENDs, a payload of less than $256$B can be inlined into the PIO 
which avoids the subsequent DMA read \cite{DBLP:conf/hpcc/MacArthurR12}.

\noindent \textbf{CPU Overhead:} We also measured the overhead (in CPU cycles) per message of different communication stacks on both the client and server. 
Again, we vary the message sizes as in the previous experiment.

Figure \ref{fig:exp:exp2b} shows that RDMA has a constant overhead on the client and the server side that is independent of the message size.
The reason is that the costs of registering a WQE on the RNIC is independent of the message size.
The actual data transfer is executed by the RNIC which acts as a co-processor to handle the given WQE.
On the client side the overhead is around $450$ cycles independent of the RDMA verb used.
The CPU overhead for atomic operations is actually the same.
Moreover, as expected, on the server side only the RECEIVE verb causes a CPU overhead.  
All other verbs that are one-sided (READ/WRITE and the atomic operations) do not cause any overhead on the server side.

The overhead of IPoIB is very different from that of RDMA.
In fact, it is much more similar to the overhead of the classical Ethernet-based TCP/IP stack (IBoEth).
The major difference to RDMA is that for IPoEth and IPoIB the per message overhead actually grows linearly with the message size once the message size exceeds the TCP window size (which was the default value of $1488$B for IPoEth and  $21888$B for IPoIB in our experiment).
Even more interesting is that for small message sizes, the per message overhead of IPoIB is even higher than for IPoEth.
For example, an $8$B message needs $7544$ cycles for IPoEth and $13264$ cycles for IPoIB.




%% file: architecture.tex
\section{Rethinking the Architecture}
\label{sec:architecture}
In this section, we discuss why the traditional architecture for distributed in-memory DBMSs is not optimal for many real-world workloads and then present novel alternatives for fast RDMA-enabled networks.
We then discuss research challenges that arise for these new architectures.

\vspace*{-3pt}
\subsection{Architectures for Fast Networks}
\label{sec:architecture:alternatives}

\vspace*{-3pt}
\subsubsection{The Traditional Shared-Nothing Architecture}
Figure \ref{fig:arch1} shows the classical shared-nothing (SN) architecture for distributed in-memory data\-bases over slow networks.
Here, the database state is partitioned over the main memory (RAM) of multiple nodes where each node has only direct access to the database partition located in its local RAM.
Furthermore, in order to implement distributed control-flow and data-flow, nodes communicate with each other using socket-based send/receive operations.

Efficient distributed query and transaction processing requires that the main goal is to maximize data-locality for a given workload by applying locality-aware partitioning schemes or by leveraging communication avoiding strategies (e.g., semi-joins).
Ideally, no communication happens between the nodes.
For many real-world workloads, however, network communication cannot be entirely avoided, resulting in large performance penalties for slow networks.
For example, even resorting to the best techniques for co-partitioning the tables \cite{Gamma:TKDE:1990,HStore:SIGMOD:2012}, it is not always possible to avoid expensive distributed join operations or distributed transactions, causing high communication costs \cite{trackj}. 
Furthermore, workloads change over time, which makes it even harder to find a good static partitioning scheme \cite{squall15}, and dynamic strategies might require moving huge amounts of data, further restricting the bandwidth for the actual work.
As a result, the network not only limits the throughput of the system, but also its scalability; the more machines are added, the more of a bottleneck the network becomes.

\vspace*{-3pt}
\subsubsection{The Shared-Nothing Architecture for IPoIB}
\begin{figure}
\begin{center}
\subfigure[SN (IPoEth)]{
   \includegraphics[width=.23\textwidth]{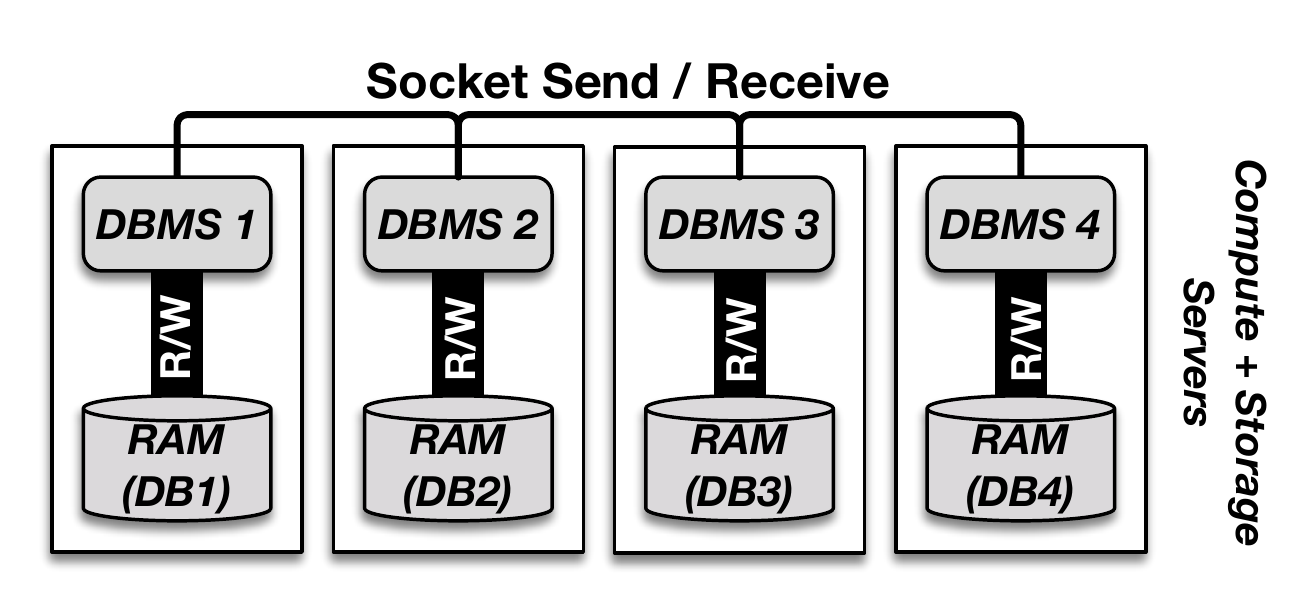}
   \label{fig:arch1}
 }
 \subfigure[SN (IPoIB)]{
   \hspace{-4ex}
   \includegraphics[width=.23\textwidth]{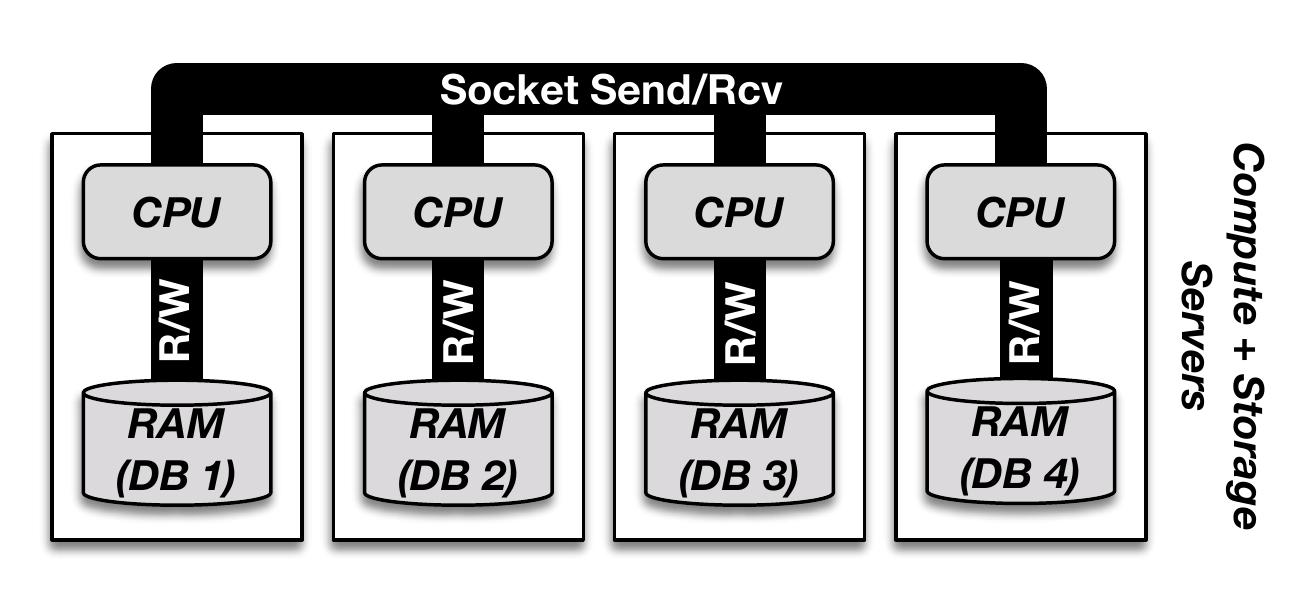}
   \label{fig:arch2}
}
\subfigure[SM (RDMA)]{
    \includegraphics[width=.23\textwidth, trim = 0mm -20mm 0mm 0mm]{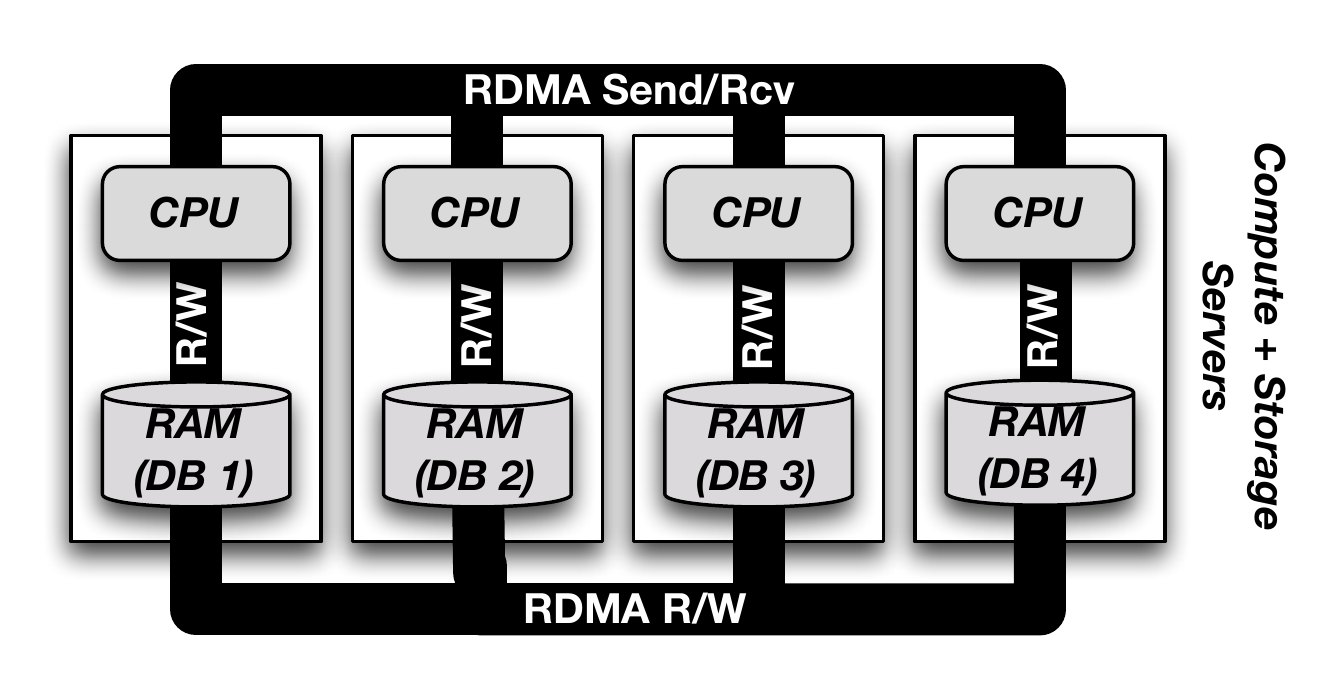}
    \label{fig:arch2-2}
}
\subfigure[NAM (RDMA)]{
   \hspace{-4ex}
   \includegraphics[width=.23\textwidth]{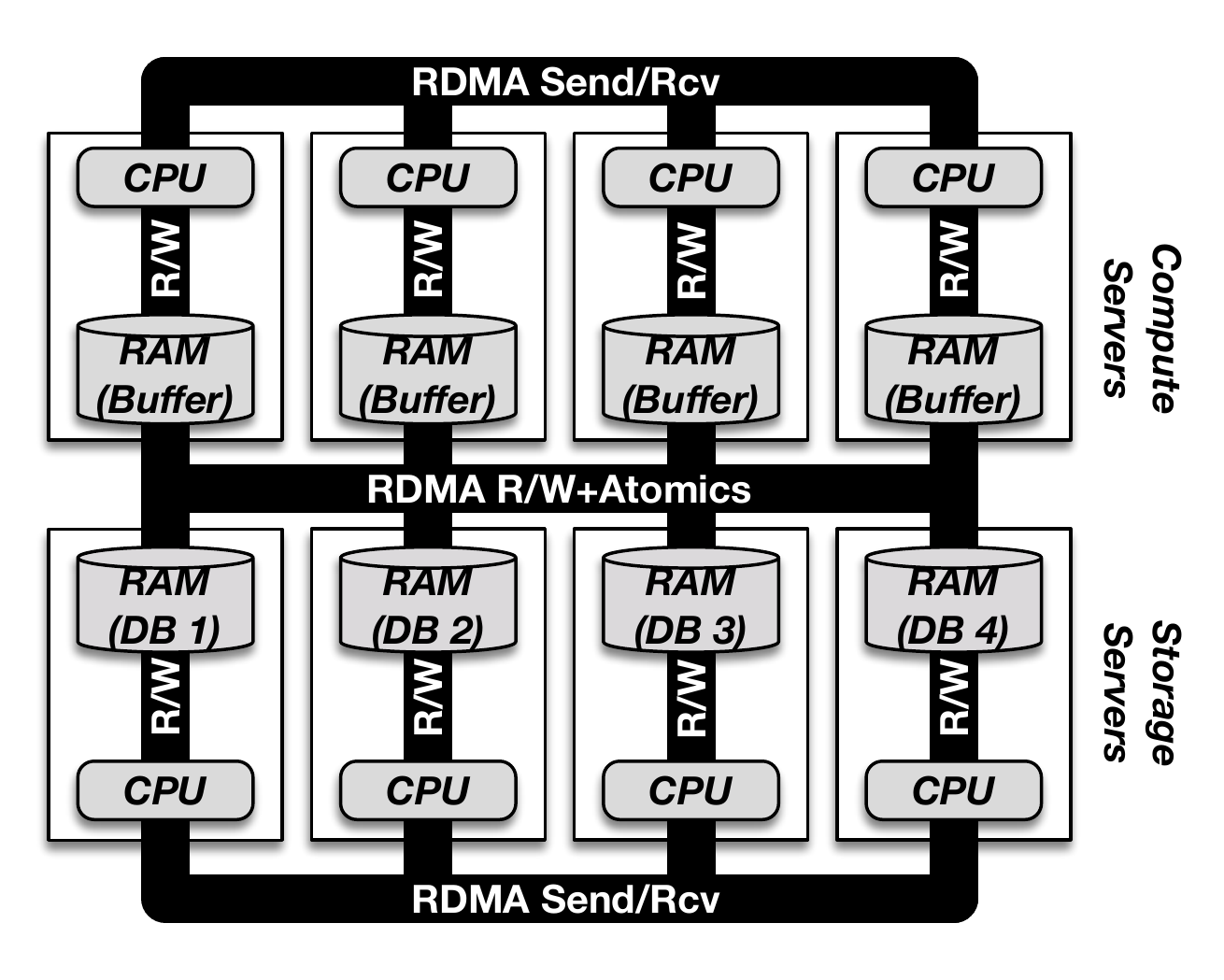}
   \label{fig:arch3}
}
\vspace*{-15pt}
\caption{In-Memory Distributed Architectures}
\label{fig:arch}
\end{center}
\vspace*{-28pt}
\end{figure}

An easy way to migrate from the traditional shared-nothing architecture to fast networks, such as InfiniBand, is to simply use IPoIB as shown in Figure~\ref{fig:arch2}.
The advantage of this architecture is that it requires almost no change of the database system itself while still benefiting from the extra bandwidth. 
In particular, data-flow operations that send large messages (e.g., data re-partitioning) will benefit tremendously from this change. 
However, as shown in Section \ref{sec:background}, IPoIB cannot fully leverage the network. 
Perhaps surprisingly, for some types of operations, upgrading the network with IPoIB can actually decrease the performance. 
This is particularly true for control-flow operations, which require sending many of small messages.
Figure~\ref{fig:exp:exp2b} shows that the CPU overhead of IPoIB is above the overhead of IPoEth for small messages. 
In fact, as we will show in Section \ref{sec:oltp}, these small differences can have a negative impact on the overall performance of distributed transaction processing. 
 
\vspace*{-3pt}
\subsubsection{The Distributed Shared-Memory Architecture}

Obviously, to better leverage the network we have to take advantage of RDMA. 
RDMA not only allows the system to fully utilize the bandwidth (see Figure~\ref{fig:exp2bw}), but also reduces network latency and CPU overhead (see Figures \ref{fig:exp2lat} and  \ref{fig:exp:exp2b}). 
Unfortunately, changing an application from a socket-based message passing interface to RDMA verbs is not trivial. 
One possibility is to treat the cluster as a shared-memory system (shown in Figure~\ref{fig:arch2-2}) with two types of communication patterns: 
message passing using RDMA-based SEND/RECEIVE verbs and remote direct memory access through one-sided RDMA READ/WRITE verbs.  

However, as stated before, there is no cache-coherence protocol.
Moreover, machines need to carefully declare the sharable memory regions a priori and connect them via queue pairs.
The latter, if not used carefully, can also have a negative effect on the performance \cite{RDMA_KV}.
In addition, a memory access via RDMA is very different than those of a shared-memory system.
While a local memory access only keeps one copy of the data around (i.e., conceptually it moves the data from main memory to the cache of a CPU), a remote memory access creates a fully-independent copy. 
This has a range of implications from garbage collection, over cache/buffer management, up to consistency protocols. 

Thus, in order to achieve the appearance of a shared-memory system, the software stack has to hide the differences and provide a real distributed shared-memory space. 
There have been recent attempts to create a distributed shared-memory architecture over RDMA \cite{farm14}. 
However, we believe that a single abstraction for local and remote memory is the wrong approach. 
Databases usually want to have full control over the memory management and because virtual memory management can get in the way of any database system, we believe the same is true for shared-memory over RDMA. 
While we had the ambitions to validate this assumption throughout our experiment, we found only one commercial offering for IBM mainframes \cite{ibm:sm}.
Instead, for our OLTP comparison, we implemented a simplified version of this architecture by essentially using a SN architecture and replacing socket communication with two-sided RDMA verbs (send and receive). We omit this architecture entirely for our OLAP comparison since two-sided RDMA verbs would have additionally added synchronization overhead to our system (i.e., an RDMA RECEIVE must be issued strictly before the RDMA SEND arrives at the RNIC), which would have simply slowed down the execution of our OLAP algorithms when compared to their NAM alternatives.

\vspace*{-3pt}
\subsubsection{The Network-Attached-Memory Architecture}
Based on the previous considerations, we envision a new type of architecture, referred to as network-attached memory  (or NAM for short) shown in Figure~\ref{fig:arch3}. 
In a NAM architecture, compute and storage are logically decoupled.
The storage servers provide a shared distributed memory pool, which can be accessed from any compute node.
However, the storage nodes are not aware of any database specific operations (e.g., joins or consistency protocols). 
These are implemented by the compute nodes. 

This logical separation helps to control the complexity and makes the system aware of the different types of main memory.
Moreover, the storage nodes can take care of issues like garbage collection, data-reorganization or metadata management to find the appropriate remote-memory address of a data page. 
Note, that it is also still possible to physically co-locate storage nodes and compute nodes on the same machine to further improve performance. 
However, in contrast to the previous architecture, the system gains more control over what data is copied and how copies are synchronized.

The NAM architecture has also several other advantages.
Most importantly, storage nodes can be scaled independently of compute nodes. 
Furthermore, the NAM architecture can efficiently handle data imbalance since any node can access any remote partition without the need to re-distribute the data before.
It should be noted that this separation of compute and storage is not new.
However, similar existing systems all use an extended key/value like interface for the storage nodes \cite{BDS3,deuteronomy15,shared-database:sigmod2015} or are focused on the cloud \cite{azure,snowflake}, instead of being built from scratch to leverage high performance networks like InfiniBand.
Instead, we argue that the storage servers in the NAM architecture should expose an interface that supports fine-grained byte-level memory access that preserves some of the underlying hardware features. 
For example, in Section \ref{sec:oltp} we show how the fine address-ability allows us to efficiently implement concurrency control. In the future, we plan to take advantage of the fact that messages arrive in order between queue pairs.  

\vspace*{-5pt}
\subsection{Challenges and Opportunities}
\label{sec:architecture:challenges}
Unfortunately, moving from a shared-nothing or shared-memory system to a NAM architecture requires a redesign of the entire distributed database architecture from storage management to query processing and transaction management up to query compilation and metadata management.

\noindent
\textbf{Transactions \& Query Processing:} 
Distributed query processing is typically implemented using a data-parallel execution scheme that leverages re-partitioning operators which shuffle data over the network. 
However, re-partitioning operators do not typically pay much attention to efficiently leveraging the CPU caches of individual machines in the cluster.
Thus, we believe that there is a need for parallel cache-aware algorithms for query operators over RDMA. 

Similarly, we require new query optimization techniques for distributed in-memory database system with high-band\-width network.
As previously mentioned, existing distributed database systems assume that the network is the dominant bottleneck. 
Therefore existing cost-models for distributed query optimization often consider the network cost as the only cost-factor \cite{Ozsu:2007:PDD:1534678}.
With fast networks and thus, a more balanced system, the optimizer needs to consider more factors since bottlenecks can shift from one component (e.g., CPU) to another (e.g., memory-bandwidth) \cite{vldbtupleware}.

Additionally, we believe that a NAM architecture requires new load-balancing schemes that implement ideas suggested for work-stealing on single-node machines \cite{morsel14}.
For example, query operations could access a central data structure (i.e., a work queue) via one-sided RDMA verbs, which contains pointers to small portions of data to be processed by a given query.
When a node is idle, it could pull data from the work queue.
That way, distributed load balancing schemes can be efficiently implemented in a decentralized manner. 
Compared to existing distributed load balancing schemes, this avoids single bottlenecks and would allow greater scalability while also avoiding stragglers. 

\noindent
\textbf{Storage Management:} 
Since the latency of one-sided RDMA verbs (i.e., read and write) to access remote database partitions is still much higher than for local memory accesses, we need to optimize the storage layer of a distributed database to minimize this latency.

One idea in this direction is to develop \emph{complex} storage access operations that combine different storage primitives in order to effectively minimize network roundtrips between compute and storage nodes.
This is in contrast to existing storage managers which offer only \emph{simple} read/write operations.
For example, in Section \ref{sec:oltp} we present a complex storage operation for a distributed SI protocol that combines the locking and validation of the 2PC commit phase using a single RDMA atomic operation.
However, for such complex operations, the memory layout must be carefully developed.
Our current prototype therefore combines the lock information and the value into a single memory location.

Modern RNICs, such as the Connect X4 Pro, provide a programmable device (e.g., an FPGA) on the RNIC.
Thus, another idea to reduce storage access latencies is to implement complex storage operations that cannot be easily mapped to existing RDMA verbs in hardware.
For example, writing data directly into a remote hash table of a storage node could be implemented completely on the RNICs in a single roundtrip without involving the CPUs of the storage nodes, hence allowing for new distributed join operations.

Finally, we believe that novel techniques must be developed that allow efficient pre-fetching using RDMA.
The idea is that the storage manager issues RDMA requests (e.g., RDMA READs) for memory regions that are likely to be accessed next and the RNIC processes them asynchronously in the background.
Moreover, the RDMA storage manager thus first polls the completion queue once a requests for a remote memory address shall be executed to check if the remote memory has already been prefetched.
While this is straightforward for sequential scanning of table partitions, index structures, which often rely on random access, require a more careful design.

\noindent
\textbf{Metadata Management and Query Compilation:} Typically, a distributed DBMS architecture has one central node which is responsible for metadata management and query compilation.
In a classical architecture this central node can either fail or become a bottleneck under heavy loads.
In a NAM architecture where all nodes can access central data structures using remote memory accesses, any node can read and update the metadata. 
Therefore, any node can compile queries and coordinate their execution. 
Thus, query compilation and metadata management exists as neither a bottleneck nor a single point of failure.


%% file: oltp.tex
\section{The Case for OLTP}
\label{sec:oltp}
The traditional wisdom is that distributed transactions, particularly when using 2-phase commit (2PC), do not scale \cite{thecasefordeterministic,h-store,onesize,Schism:VLDB:2010, Andy-thesis,locking12}. 
In this section, we show that this is the case on a shared-nothing architecture over slow networks and then present a novel protocol for the NAM architecture that can take full advantage of the network and, theoretically, removes the scalability limit. 

\vspace*{-3pt}
\subsection{Why 2PC does not scale}
\label{sec:oltp:notscalable}
In this section, we discuss factors that hinder the scalability of distributed transaction processing over slow networks. 
Modern DBMSs employ Snapshot Isolation (SI) to implement concurrency control and isolation because it promises superior query performance compared to lock-based alternatives.  
The discussion in this section is based on a 2PC protocol for generalized SI~\cite{SI-middleware, GeneralizedSI}. 
However, the findings can also be generalized to more traditional 2PC protocols~\cite{classic-PC}.

\subsubsection{Dissecting 2PC}
\label{sec:oltp:2pc}
\begin{figure}
\begin{center}
\subfigure[Traditional SI]{
   \includegraphics[height=4.3cm]{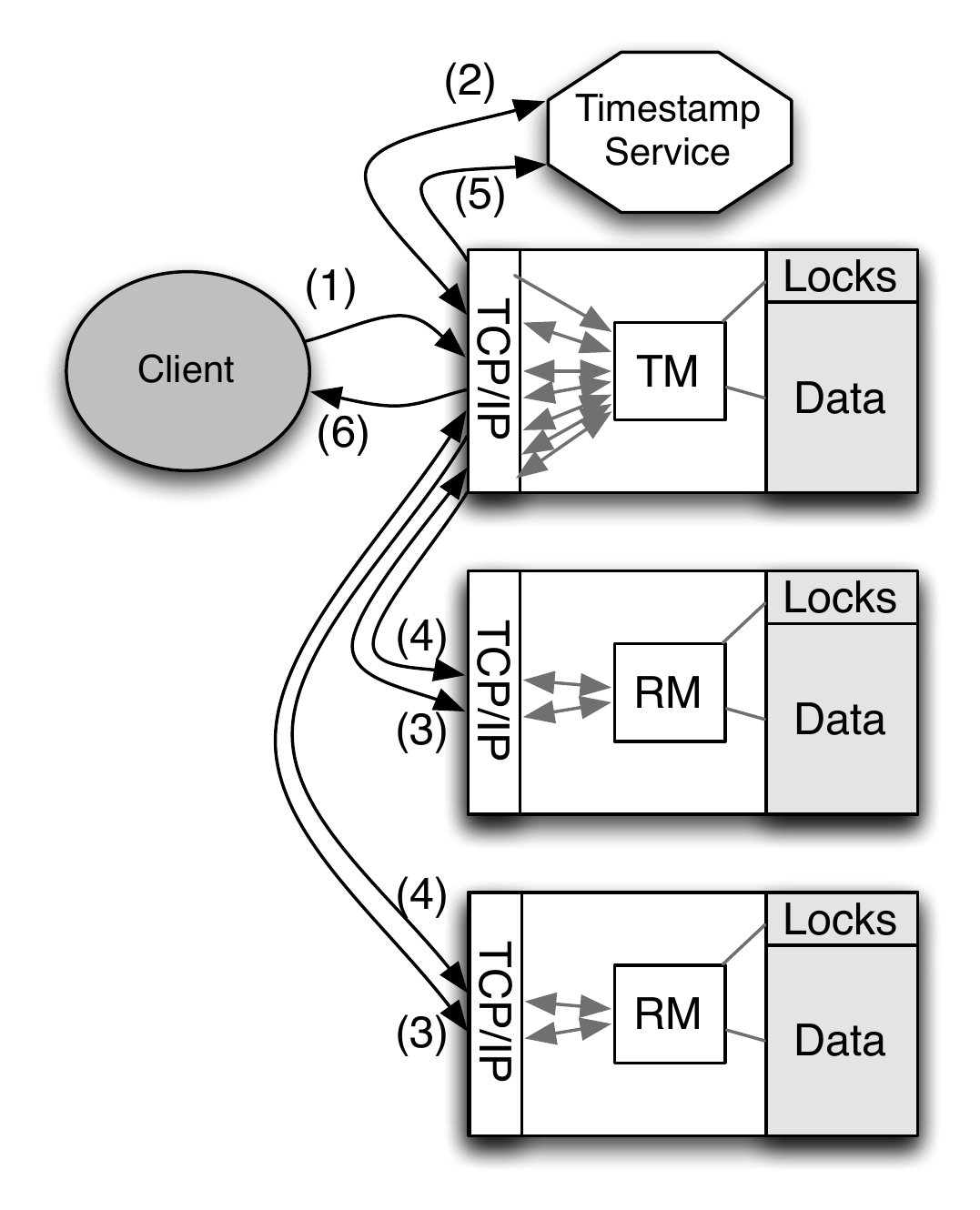}
   \label{fig:old-commit}
 }
 \subfigure[RSI Protocol]{
   \includegraphics[height=4.3cm]{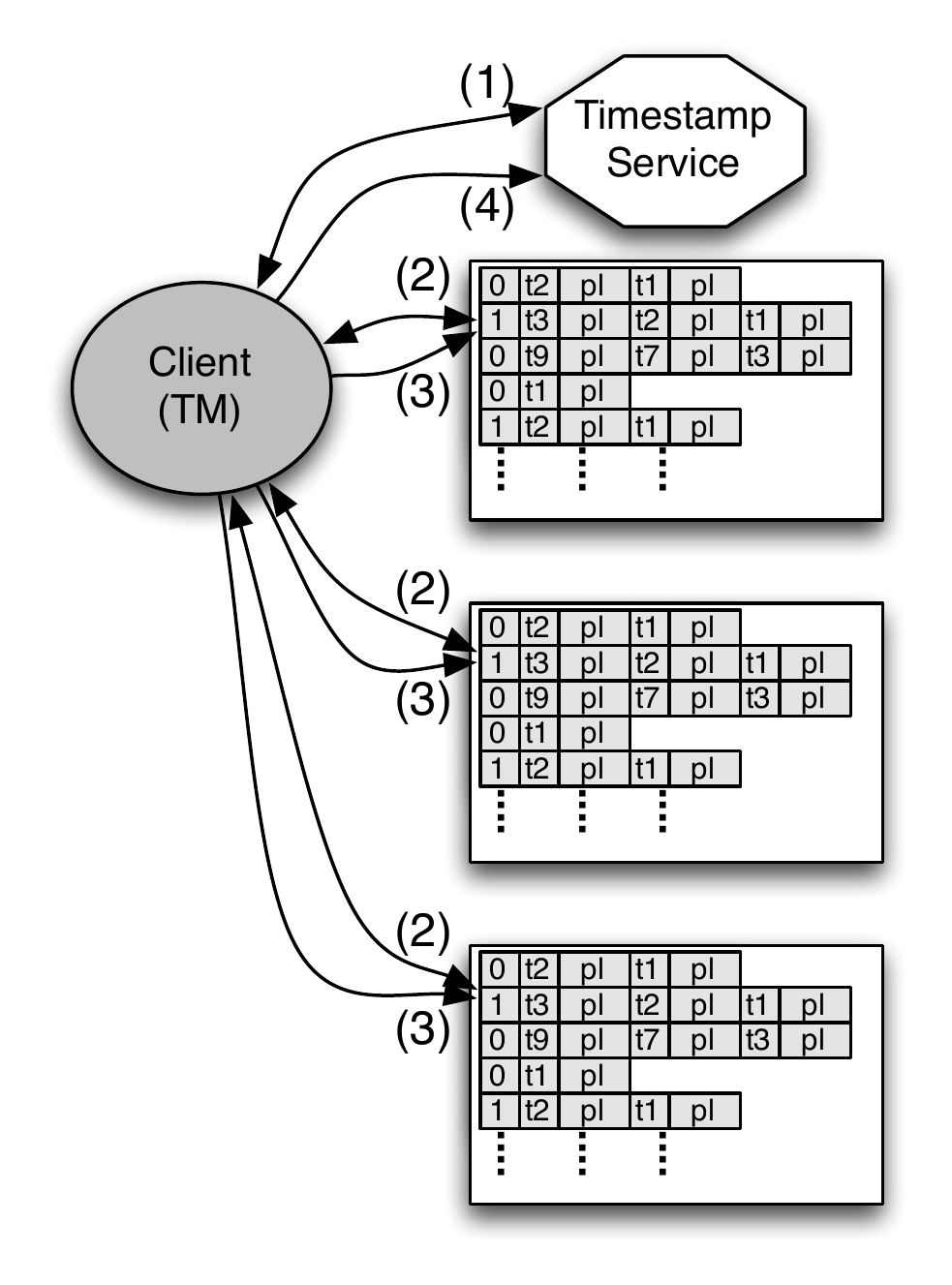}
   \label{fig:new-commit}
}
\vspace{-12pt}
\caption{Distributed 2PC Commit Protocols for SI}
\vspace{-25pt}
\label{fig:commit}
\end{center}
\end{figure}

Figure~\ref{fig:old-commit} shows a traditional (simplified) protocol using 2-phase commit with generalized SI guarantees \cite{SI-middleware, GeneralizedSI}, assuming that the data is partitioned across the nodes (i.e., shared-nothing architecture)  and without considering the read-phase (see also \cite{BDS3,Spanner,Oracle-RAC}). 
That is, we assume that the client (e.g., application server) has read all necessary records to issue the full transaction using a (potentially older) read-timestamp (RID), which guarantees a consistent view of the data.
After the client finishes reading the records, it sends the commit request to the transaction manager (TM) [one-way message 1].
While Figure~\ref{fig:old-commit} only shows one TM, there can be more, evenly distributing the load across nodes. 

As a next step, the TM requests a {\em commit timestamp} ({\em CID}) [round-trip message 2].
In this paper, we assume that globally ordered timestamps are given out by an external service, as suggested in \cite{BDS3} or \cite{Spanner}.
Since the timestamp service implementation is orthogonal, we simply assume that the timestamp service is not a potential bottleneck when using approaches like Spanner \cite{Spanner} or epoch-based SI \cite{silo}.

After the TM received the CID, it prepares the other nodes involved in the transaction through {\em prepare} messages to the resource managers (RM) as part of 2PC [round-trip message 3]. 
Each RM (1) checks to see if the records in it's partition have been modified since they have been read by the transaction and (2) locks each tuple to prevent updates by other transactions after the validation\cite{SAP_SI}.
This normally requires checking if any of the records of the write-sets has a higher CID than the RID. 

If the TM was able to prepare all involved RMs, the transaction can be committed by sending a {\em commit} message to all RMs [round-trip message 4], which installs the new version (value and commit-timestamp) and releases the locks. 
Moreover, in order to make the new value readable by other transactions, TM needs to wait until the second phase of 2PC completes [message 4], and then inform the timestamp service that a new version was installed [one-way message 5]. 
For the remainder, we assume that the timestamp service implements a logic similar to \cite{BDS3} or Oracle RAC \cite{Oracle-RAC} to ensure the SI properties. 
That is, if a client requests an RID, the timestamp service returns the largest committed timestamp.
Finally, the TM notifies the client about the outcome of the transaction [one-way message 6].

Overall the protocol requires 9 one-way message delays if done in the previously outlined sequential order. 
However, some messages can be done in parallel: the commit-time\-stamp [message 2] can be requested in parallel to preparing the resource manager [message 3] since the commit-timestamp is not required until the 2nd phase of 2PC [message 4].
This simplification is possible since we assume blind writes are not allowed; therefore a transaction must read all data items (and their corresponding RID) in its working set before attempting to commit.  
Similarly, the client can be informed [message 6] in parallel with the 2nd phase of 2PC [message 4].
This reduces the number of message delays to 4 until the client can be informed about the outcome (one-way message 1, round-trip 3, one-way message 5), and to at least 6 until the transaction becomes visible (one-way message 1, round-trips 3 and 4, one-way message 6). 
Compared to a centralized DBMS, the 6 message delays required for 2PC substantially increases the execution time for a transation.

Unlike the described 2PC protocol, a traditional 2PC protocol~\cite{classic-PC} does not require a timestamp service.
However, a traditional 2PC protocol which consists of a {\em prepare} and a {\em commit/abort} phase still requires 6 message delays in total (including client notification).
Thus, the following discussion is not specific to SI and can be generalized towards other more traditional 2PC protocols as well. 

\vspace*{-3pt}
\subsubsection{Increased Contention Likelihood}
\label{sec:oltp:contention}

The increased transaction latencies due to message delays increase the chance of contention and aborts. 
As outlined in Section~\ref{sec:background}, the average latency for small one-way messages over Ethernet is roughly 35$\mu$s, whereas the actual work of a transaction ranges from 10- 60$\mu$s today if no disk or network is involved \footnote{For instance \cite{low-overhead} reported $64\mu s$ for a single partition transaction on an ancient 2008 Xeon processor} \cite{h-store,HanaOverview}.
That is, for short-running transactions, the dominant factor for latency is the network and 2PC just amplifies the bottleneck. 

In order to model the contention rate effect, we assume an M/M/1 queue $X$ to estimate the number of waiting, i.e., conflicting, transactions for a given record $r$ with some arrival rate $\lambda$.
With this model, a $6\times$ increase in transaction processing time  (i.e., also referred to as service time $t$) yields to a service capacity decrease of 
$\mu = 1 / (6  t)$ and thus, an increased conflict likelihood of $P(X>=0) = 1 - P(X=0)  =  1 - ( 1 - \lambda / \mu ) = 6  \lambda t$.
However, a transaction rarely consists of a single record. 
With $n$ records, the likelihood of a conflict increases to $1- \prod_n P(X=0) = 1 - (1 - 6 \lambda t) ^n$, if we employ the simplifying assumption that the access rate to all records is similar and independent. 
So the intuition that the likelihood of conflicts with 2PC increases is true. 

However, we did not consider the read-phase and it is easy to show that the relative difference is less severe as more records are read (it adds a fixed cost to both).
In addition, a redesign of the commit protocol to use RDMA verbs can significantly decrease the conflict likelihood since the latency is much lower for small messages (see also Figure~\ref{fig:exp2lat}).
Furthermore, recent work has shown that most conflicts can be avoided using commutative updates \cite{bailis2013eventual}.
In fact, using newer consistency protocols, it is even possible to take advantage of non-blocking commutative updates while preserving limit constraints (e.g., a product's stock cannot fall below 0), high availability, and using no centralized coordination \cite{MDCC}.
As a result, we believe that the increased conflict-likelihood should no longer be an argument that distributed transactions can not scale.

\begin{table*}
\center
    \tiny
    \begin{tabular}{| l | l | l | l | l | l | l |}
    \hline
     Look &  $CID_N$ & $Record_N$  & $CID_{(N-1)}$ & $Record_{(N-1)}$  & $CID_{(N-2)}$ & $Record_{(N-2)}$  \\ 
     1-Bit &  (63 Bit) &  (m Bits) & (64 Bit) &  (m Bits)  &  (64 Bit) & (m Bits) \\\hline  \hline
     0 & 20003 & ("Name1", "Address1") & & & & \\ \hline
     0 & 23401 & ("Name2", "Address2") & 22112 & ("Name2", "OldAddr")  & & \\ \hline
     1 & 24401 & ("Name3", "Address3") & 22112 & ("Old3", "Old3")  & & \\ \hline
    \end{tabular}
    \vspace*{-5pt}
    \caption{Potential Data Structure for RDMA-enabled Snapshot Isolation}
      \label{tab:datastructure} 
      \vspace*{-15pt}
\end{table*}

\subsubsection{CPU Overhead}
\label{sec:oltp:overhead}
The increased likelihood of conflicts is, however, not the main reason why distributed transactions in general, and 2PC specifically, are doomed to be non-scalable.
With an increasing number of server nodes, the number of network messages also increases.
In a centralized system, the DBMS only has to handle 2 messages per transaction (i.e., the request and response to the client). 
If we assume that the clients can be scaled independently from the server (and are not further considered), the server has to handle one receive message ($m_r$) and one send message ($m_s$) in the centralized case. 
Without RDMA, the receiver and the sender {\bf both} have to spend CPU cycles for every message. 

In our distributed scenario of Figure~\ref{fig:old-commit} with one TM server and $n$ involved RMs ($n=2$ in Figure~\ref{fig:old-commit}),  every transaction requires $m_r = 2 + 4 \cdot n $ and $m_s = 3 + 4 \cdot n$ messages.
Assuming that sends and recieves are similarly expensive we get $m = m_r + m_s = 5 + 8 \cdot n $, which is 
significantly more than the single or centralized case. 

Let us assume that a transaction always has to access all $n$ servers.
If we assume that every server has $c$ cores (each of which is able to execute $cycles_c$ per second) and a message costs $cycles_m$, then a very optimistic upper bound on the number of transactions per second is $trx_u = (c \cdot cycles_c \cdot (n+1)) / (5 + 8 \cdot n) \cdot cycles_m$.
On a modern $3$ node cluster with $2.2$GHz $8$-core CPUs and assuming $3,750$ cycles per message (see Figure~\ref{fig:exp:exp2b}), this leads to $\approx 647,000$ trx/seconds.
More interestingly, though, if we increase the cluster to $4$ nodes with the same hardware configuration the maximum throughput goes down to $634,000$. 
Of course, these are only back-of-the-envelope calculations but they show that the message overhead essentially consumes all the added CPU power, making the system inherently unscalable (if the workload cannot be partitioned). 

Message batching can help, but with increased batch sizes, the processing time per message also increases. 
Furthermore, without redesigning the protocol and data structures, the CPU overhead will remain one of the most dominant bottlenecks. 
For instance, as Figure~\ref{fig:exp:exp2a} and Figure~\ref{fig:exp:exp2b} show, the IPoIB implementation over our FDR network helps increase the bandwidth and reduce the latency compared to IPoEth, but it does not reduce the CPU overhead, and in some cases it may exacerbate the situation.


\vspace*{-8pt}
\subsubsection{Discussion}
\label{sec:oltp:bandwidth}
The traditional wisdom that distributed transactions, especially 2PC, do not scale is true.
First, distributed transactions increase the contention rate, but not significantly.
Second, the protocol itself (not considering the message overhead) is rather simple and has no significant impact on the performance (2PC simply checks if a message arrived and what it contained). 
What remains the dominating factor is the increased CPU-load for handling the messages and even the network bandwidth. 
Assuming a $10$Gb Ethernet with 3 servers, an average record size of $1$KB, and that a transaction updates on average three records, at least $3$KB have to be read and written per transaction.
This limits the total throughput to $\approx 218,500$ transactions per second. 

As a result, complicated partitioning schemes have been proposed to avoid distributed transactions as much as possible \cite{Schism:VLDB:2010,onesize, LocalityAware}.
While it is a solution, it imposes a new set of challenges for the developer and some workloads (e.g., social-graph data is notoriously hard to partition).

\vspace*{-3pt}
\subsection{RSI: An RDMA-based SI Protocol}
\label{sec:oltp:proto}
Fast high-bandwidth networks such as InfiniBand are able to lift the two most important limiting factors: CPU overhead and network bandwidth.
However, as our experiments show, the scalability is severely limited without changing the techniques themselves.
Therefore, we need to redesign distributed DBMSs for RDMA-based architectures.

In this section, we present a novel RDMA-based SI protocol, called {\em RSI} that is designed for the network-attached memory (NAM) architecture . 
We have also implemented the traditional SI protocol discussed before using two-sided RDMA verbs instead of TCP/IP sockets as a simplified shared-memory architecture.
Both implementations are included in our experimental evaluation in Section \ref{sec:oltp:eval}.
In the following, we only focus on our novel RSI protocol.

At its core, RSI moves the transaction processing logic to the client (i.e., compute nodes) and make the servers (i.e., storage nodes) ``dumb'' as their main purpose is to share their main memory to the clients.
Moreover, clients implement the transaction processing logic through one-sided RDMA operations (i.e., the client is the transaction manager) allowing any compute node to act as a client that can access data on any storage node (i.e., a server).
This design is similar to~\cite{BDS3}, but optimized for direct memory access rather than cloud services. 
Moving the logic to the client has several advantages. 
Most importantly, scale-out becomes much easier since all CPU-intensive operations are done by the clients, which are easy to add.
The throughput of the system is only limited by the number of RDMA requests that the server's RNICs (and InfiniBand switches) can handle.
Since several RNICs can be added to one machine, the architecture is highly scalable (see also Section~\ref{sec:oltp:eval}).
In addition,
(1) load-balancing is easier since transactions can be executed on any node independent of any data-locality, and 
(2) latencies are reduced as clients can fetch data directly from the servers without involving the TM. 

As before, we assume that reads already have happened and that the transaction has an assigned read timestamp, RID. 
First, the client (acting as the TM) contacts the timestamp service to receive a new commit timestamp CID.
In our implementation, we pre-assign timestamps to clients using a bitvector with 60k bits. 
The first bit in the vector belongs to client 1 and represents timestamp 1, up to client $n$ representing timestamp $n$. 
Afterwards, position $n+1$ again belongs to client 1 and so on. 
Whenever a timestamp is used by a client, it ``switches'' the bit from 0 to 1. 
With this scheme, the highest committed timestamp can be determined by finding the highest consecutive bit in the vector.
If all bits are set by a client, we allow clients to ``wrap'' and start from the beginning.
Note, that wrapping requires some additional bookkeeping to avoid that bits are overwritten.

This simple scheme allows the clients to use timestamps without having a synchronization bottleneck but implicitly assumes that all clients make progress at roughly the same rate.
If this assumption does not hold (e.g., because of stragglers, long running transactions, etc.), additional techniques are required to skip bits, which go beyond the scope of this paper.
Also note, to ensure a fair comparison, we use the same technique for the traditional protocol implementation; even though we know from our experiments that it does not provide any benefits in that case.

Next, the client has to execute the first phase of 2PC and check if the version has not changed since it was read (i.e., validation phase of 2PC). 
As before, this operation requires a lock on the record to prevent other transactions from changing the value after the validation and before the transaction is fully committed. 
In a traditional design, the server would be responsible of locking and validating the version.
In order to make this operation more efficient and ``CPU-less'', we propose a new storage layout to allow direct validation and locking with a single RDMA-operation shown in Table~\ref{tab:datastructure}. 
The key idea is to store up to $n$ versions of a fixed-size record of $m$-bits length in a fixed-size {\em slotted} memory record, called a {\bf ``record block''}, and have a global dictionary (e.g., using a DHT) to exactly determine the memory location of any record within the cluster. 
We will explain the global dictionary and how we handle inserts in the next subsections and assume for the moment, that after the read phase all memory locations are already known. 
How many slots (i.e., versions) a record block should hold depends on the update and read patterns as it can heavily influence the performance. 
For the moment, assume that every record has $n$ = $max(16 KB$ / $\textit{record-size},2)$ slots for different record versions and that every read retrieves all $n$ slots.
From Figure~\ref{fig:exp2lat} we know that transferring $1$KB to roughly $16$KB makes no difference in the latency threfore making $n$ any smaller has essentially no benefit.
Still, for simplicity, our current implementation uses $n=1$ and aborts all transactions which require an older snapshot. 

The structure of a slot in memory is organized as follows: the first bit is used as a lock (0=no-lock, 1=locked) while the next $63$ bits contain the latest commit-id (CID) of the most recent committed record, followed by the payload of the record, followed by the second latest CID and payload and so on, up to $n$ records. 
Using this data structure, the TM (i.e., the client) is directly able to validate and lock a record for a write using a {\em compare-and-swap} operation on the first 64 bits [round-trip message 2]. 
For example, assume that the client has used the RID $20003$ to read the record at memory address 1F (e.g., the first row in Table~\ref{tab:datastructure}) and wants to install a new version with CID $30000$.
A simple RDMA {\em compare-and-swap} operation on the first $64$ Bits of the record at address $1F$ with test-value $20003$, setting it to  {$1 << 63 | 20003$)}, would only acquire the lock if the record has not changed since it was read by the transaction, and fails otherwise. 
Thus, the operation validates and prepares the resource for the new update in a single round-trip. 
The TM uses the same technique to prepare all involved records (with SI inserts always succeeding).

\begin{figure}
\begin{center}
\includegraphics[width=.3\textwidth]{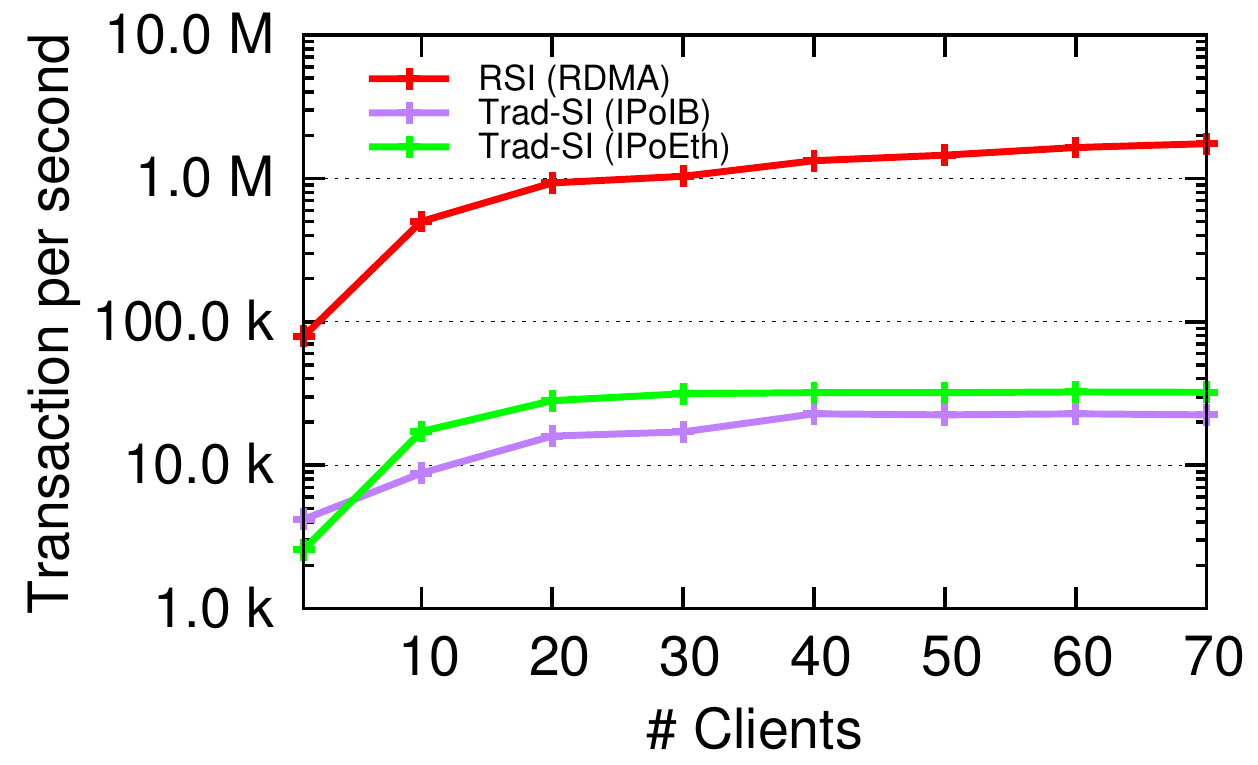}
\vspace{-12pt}
\caption{RSI vs 2PC}
\vspace{-29pt}
\label{fig:scale}
\end{center}
\end{figure}

If the {\em compare-and-swap} succeeds for all intended updates of the transaction, the transaction is guaranteed to be successful and the TM can install a new version. 
The TM therefore checks if the record block has a free slot, and, if yes, inserts its new version at the head of the block and shifts the other versions to the left.
Afterwards, the TM writes the entire record block with a signaled WRITE to the memory location of the server [message 3]. 

Finally, when all the writes have been successful, the TM informs the timestamp service about the outcome [message 3] as in the traditional protocol. This message can be sent unsignaled. 
Overall, our RDMA-enable SI protocol and storage layout requires 3 round-trip messages and one un\-signaled message, and does not involve the CPU in the normal operational case.
As our experiments in the next section will show, this design enables new dimensions of scalability.


\vspace*{-3pt}
\subsection{Experimental Evaluation}
\label{sec:oltp:eval}
To evaluate the algorithms, we implemented the traditional SI protocol (Figure~\ref{fig:old-commit}) on the shared-nothing architecture with IPoETH (Figure~\ref{fig:arch1}) and IPoIB (Figure~\ref{fig:arch2}).
We also implemented a simplified variant of the shared-memory architecture (Figure~\ref{fig:arch2-2}) by replacing TCP/IP sockets with two-sided RDMA verbs (requiring significantly modifiying memory management). 
We slightly adjusted the traditional SI implementation by using a local time-stamp server instead of a remote service (i.e., we gave the traditional implementation an advantage). 
Finally, our RSI protocol implements the NAM architecture  (Figure~\ref{fig:arch3}) and uses an external timestamp service as described earlier.

We evaluated all protocols on an $8$-node cluster using the same configuration as in Section~\ref{sec:background:micro}.
We use four machines to execute the clients, three as the NAM storage-servers, and one as the timestamp server (or as the transaction manager in traditional).
We measured both protocols with a simple and extremely write-heavy workload, similar to the checkout transaction of the TPC-W benchmark. 
Every transaction reads $3$ products, creates $1$ order and $3$ orderline records, and updates the stock of the products.
As base data, we created 1 million products (every record is roughly 1KB) to avoid contention, and all data was evenly distributed across the machines. 
Clients wait until a transaction is finished before issuing the next transaction.

Figure~\ref{fig:scale} shows the scalability of the traditional SI-protocol and our new RSI protocol with the number of client threads varied from $1$ to $70$.
The traditional SI-protocol over IPoIB has the worst scalability, with $\approx 22,000$ transactions per second, whereas IPoEth achieves $\approx 32,000$ transactions per second.
The IPoIB implementation performs worse because of the less efficient TCP/IP implementation for IPoIB, which plays an important role for small messages.
In contrast, our RSI protocol achieved a stunning $\approx 1.8$ million {\em distributed} transactions per second.
The shared-memory architecture using two-sided RDMA verbs achieved a throughput of $1.1$ million transaction per second, or only $66\%$ of our RSI protocol (line omitted due to overlap with RSI because of the log-scale). 
However, we also noticed that the two-sided RDMA verb implementation not only stops scaling after 40 clients, but that the throughput also decreases to only $\approx 320,000$ transaction per second with 70 clients, while our RSI implementation scales almost linearly with up to 60 clients.
One reason for the decrease in performance is that the transaction managers become one of the major bottlenecks. 
However, our RSI implementation no longer scaled linearly after 60 clients, since we only had one dual-port FDR 4x RNIC per machine, with a total bandwidth of 13.8GB/s. 
With the three 1KB records per transactions, we can achieve a theoretical maximum throughput of $\approx 2.4M$ transactions per second (every transaction reads/writes at least 3KB).
For greater than 60 clients, the network is saturated.

We therefore speculate that distributed transactions no longer have to be a scalability limit when the network bandwidth matches the memory bandwidth and that complex partitioning schemes might become obsolete in many scenarios (note, that they can still reduce latency and/or to manage hot items).

%% file: olap.tex
\section{The Case for OLAP}
\label{sec:olap}
In order to motivate the redesign of distributed DBMSs for OLAP workloads, we first discuss why existing distributed algorithms, which were designed for a shared-nothing architecture over slow networks, are not optimal for fast RDMA-capable networks.
Then, we present novel RDMA-optimized operators for the NAM architecture, which require fundamental redesigns of central components (e.g., memory management, optimization), as discussed in Section~\ref{sec:architecture:challenges}.
This paper focuses on distributed joins and aggregates, which are the predominant operators in almost any OLAP workload.

\vspace*{-0pt}
\subsection{Existing Distributed OLAP Operators}
\label{sec:olap:existing}
The most network-intensive operation in OLAP workloads is the distributed join~\cite{roediger}.
Most distributed join algorithms have three components: (1) a local join algorithm, (2) a  partitioning scheme, and (3) an optional reduction technique.
All three components can be combined in different ways. 
For example, either a hash or sort-merge join could be used as the local join algorithm, whereas partitioning schemes range from static to dynamic hash partitioning~\cite{Gamma:TKDE:1990}. 
Similarly, several techniques to reduce the partitioning cost have been proposed, the most prominent being a semi-join reduction using a Bloom filter~\cite{bloom}. 

The following section explains the most common partitioning technique for distributed join algorithms over shared-nothing architectures, the grace hash join (GHJ), in more detail.
Later, we expand the distributed join algorithm with an additional semi-join reduction using Bloom filters to further reduce communication. 
For both, we develop a simple cost model and argue why these algorithms are (in most cases) not optimal for in-memory databases over fast RDMA-capable networks.
Throughout the rest of this section, we assume that there is no skew in the data (i.e., before and after partitioning all nodes hold the same data).

\vspace*{-3pt}
\subsubsection{An Optimized Grace Hash Join}
\label{sec:olap:grace}
The GHJ executes a distributed join in two phases. 
In the first phase (partitioning phase), the GHJ scans the input relations and hash-partitions them on their join key such that the resulting sub-relations can be joined in the second phase locally per node.
The cost of the GHJ $T_{GHJ}$ is therefore given by the sum of the runtime of the partitioning phase $T_{part}$ and the local join phase $T_{join}$.


We do not consider any static pre-partitioning, so the cost for repartitioning can be split into the cost of partitioning the two join relations $R$ and $S$.
The cost of repartitioning $R$ can now further be split into the cost of (1) reading the data on the sender, (2) transferring the data over the network, and (3) materializing the data on the receiver.
Assuming that the cost of sending $R$ over the network is $T_{net}(R)= w_r \cdot |R| \cdot c_{net}$ and scanning $R$ in-memory is $T_{mem}(R)= w_r \cdot |R| \cdot c_{mem}$, with $|R|$ being the number of tuples, $w_R$ being the width of a tuple $r \in R$ in bytes, and $c_{net}$ ($c_{mem}$) the cost of accessing a byte over the network (memory), the repartitioning cost of $R$ can be expressed as:

\vspace*{-1.5ex}
\begin{equation*}\small
\begin{split}
T_{part}(R) & =\underbrace{T_{mem}(R)}_\text{Reading (sender)} + \underbrace{T_{net}(R)}_\text{Shuffling (net)} + \underbrace{T_{mem}(R)}_\text{Writing (receiver)} \\
&=  w_r \cdot |R| \cdot c_{mem} + w_r \cdot |R| \cdot c_{net} + w_r \cdot |R| \cdot c_{mem} \\
&= 2 \cdot w_r  (\cdot c_{mem} \cdot |R|  + c_{net} \cdot |S|)
\end{split}
\label{eq:grace:part}
\end{equation*}

The partition cost for $S$ is similar. 
Note that we ignore any CPU cost, as we assume that the limiting factor is the memory and network access (not the CPU), which is reasonable for a simple hash-based partitioning scheme. 

For the local join algorithm of the GHJ, we use the fastest local in-memory join algorithm, the (parallel) radix join~\cite{mmjoins:paper}.
The radix join proceeds in two phases. 
In the first phase, the radix join scans each input relation and partitions the relations locally into cache-sized blocks using multiple passes over the data. 
As shown in \cite{mmjoins:paper}, with software managed buffers, most relations can efficiently be partitioned with one pass. 
After partitioning the data, the radix join scans the relations again to join the cache-sized blocks.
Existing work (\cite{mmpart} and \cite{mmjoins:paper}) has shown that both phases of the radix join are memory-bandwidth bound.
Thus, we can estimate the total cost for the local radix join as:

\vspace*{-1.5ex}
\begin{equation*}\small
\begin{split}
T_{join}(R, S) &=\underbrace{ (T_{mem}(R) + T_{mem}(S)) }_\text{Radix Phase 1}+\underbrace{ ( T_{mem}(R) + T_{mem}(S) )}_\text{Radix Phase 2} \\
&= 2 \cdot c_{mem}    \cdot  ( w_r \cdot |R| +  w_s \cdot |S|)
\end{split}
\label{eq:grace:join}
\end{equation*}

The total runtime of the GHJ $T_{GHJ}$ is therefore:

\vspace*{-1.5ex}
\begin{equation*}\small
\begin{split}
T_{GHJ} &= T_{part}(R) + T_{part}(S) + T_{join}(R,S) \\
&= (w_r |R| + w_s |S|) \cdot  (4  \cdot c_{mem} + c_{net})
\end{split}
\label{eq:grace:final}
\end{equation*}

\vspace*{-3pt}
\subsubsection{Adding Semi-Reduction using Bloom Filters}
\label{sec:olap:bloom}
As shown in the final cost equation from the previous section, the GHJ requires roughly four times more memory accesses than network transfers. 
However, in distributed in-memory DBMSs, the network cost typically dominates up to $90\%$ of the runtime of a join~\cite{roediger}.
Thus, state-of-the-art join algorithms (e.g., track join~\cite{trackj}, Neo-Join~\cite{roediger}) try to reduce network traffic through cost-intensive computations (e.g., Neo-Join uses a linear solver) or multiple communication round-trips to partition the data to further optimize the network traffic.

Here, we focus on the most traditional approach: a semi-join reduction using a Bloom filter.
The core idea of the semi-join reduction is to send only tuples in the input relations $R$ and $S$ that have a join partner in the other relation.
Therefore, the algorithm first creates Bloom filters $b_R$ and $b_S$ over the join keys of $R$ and $S$, respectively.
Then, $b_R$ and $b_S$ are copied across all nodes that hold a partition of $S$ and $R$, respectively, and each node uses its Bloom filter to filter out the tuples that are guaranteed to have no join partner (i.e., if the Bloom filter matches a join key, it must be sent). 

The cost of creating $b_R$ includes both a scan over the data $T_{mem}(R)$ and transmission over the network $T_{net}(b_R)$:

\vspace*{-1.5ex}
\begin{equation*}\small
T_{bloom}(R) = \underbrace{T_{mem}(R)}_\text{Create Reducer} + \underbrace{T_{net}(b_R)}_\text{Ship Reducer}
\label{eq:bloom:reducer}
\end{equation*}

However, the size of the Bloom filter $b_r$ is normally very small, so that $T_{bloom}(R)$ can be disregarded. 
Assuming that $sel_S(b_R)$ is the selectivity of the Bloom filter $b_R$ over relation $S$ (including the error rate of the Bloom filter), the total cost for a GHJ with a semi-join reduction using Bloom filters is:

\vspace*{-1.5ex}
\begin{equation*}\small
\begin{split}
T_{ghj+bloom} =& \underbrace{T_{bloom}(R) + T_{bloom}(S)}_\text{Create Bloom-Filter} + \\
 & \underbrace{T_{part}(sel_R(b_S) \cdot R) + T_{part} (sel_S(b_R) \cdot S)}_\text{Reduced Partitioning Cost} + \\
 & \underbrace{T_{join}(sel_R(b_S) \cdot R, sel_R(b_R) \cdot S)}_\text{Reduced Join Cost}
\end{split}
\end{equation*}

This equation models the cost of creating the Bloom filter plus the reduced partitioning and join costs. 
Assuming that the selectivity between both relations is the same, $sel=sel_R(b_S)=sel_S(b_R)$ leads to this simplified total cost:

\vspace*{-1.5ex}
\begin{equation*}\small
\begin{split}
T_{join+bloom} =&  (w_r |R| + w_s |S|) \cdot \\ 
&( c_{mem}  + 4 \cdot sel \cdot c_{mem} + sel \cdot c_{net})
\label{eq:bloom:final}
\end{split}
\end{equation*}

\vspace*{-3pt}
\subsubsection{Discussion}
\begin{figure}
\begin{center}
\subfigure[IPoEth]{
   \hspace{-7ex}
   \includegraphics[width=.25\textwidth]{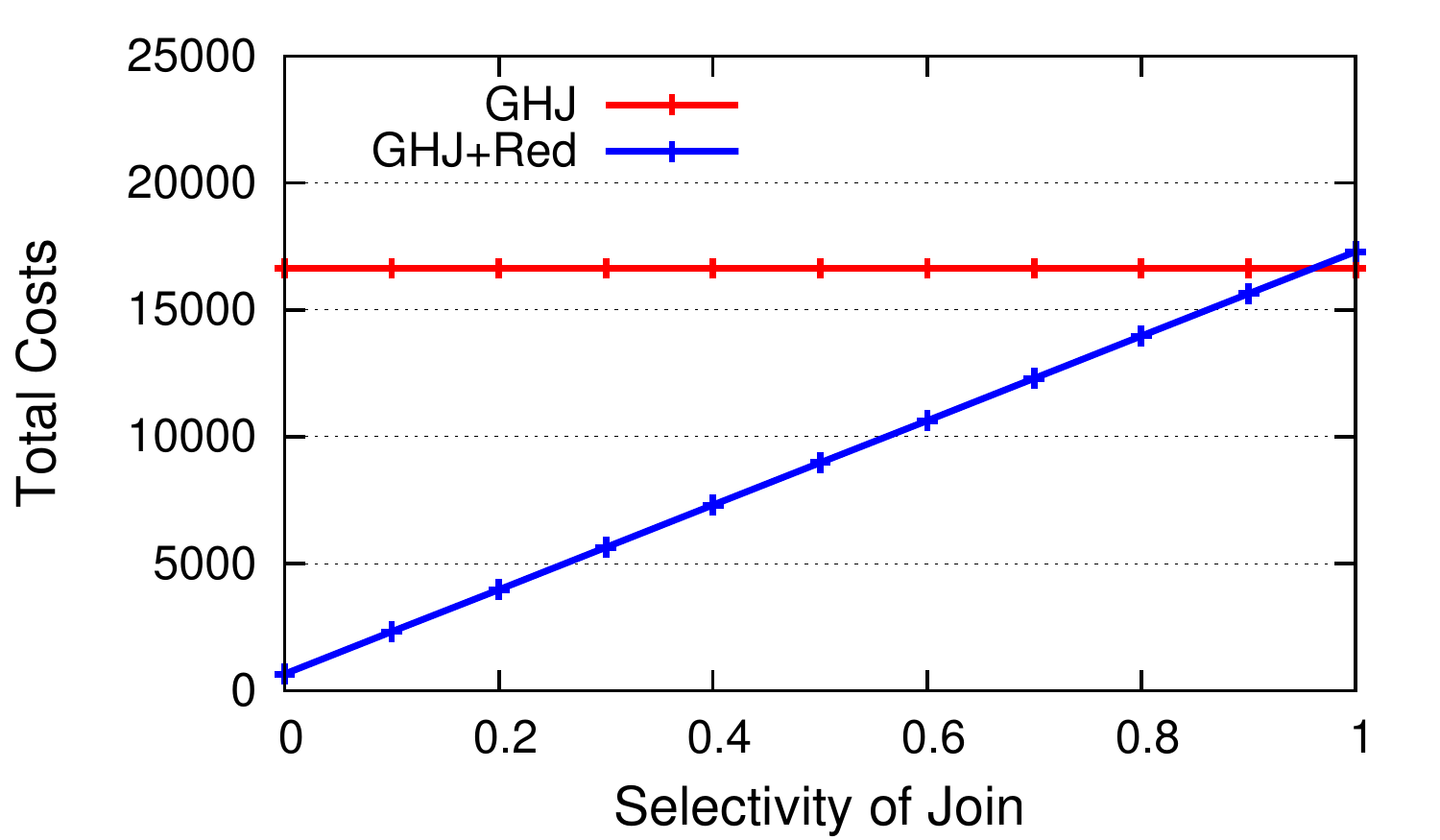}
   \label{fig:olap:cost_slow}
 }
 \subfigure[IPoIB and RDMA]{
   \hspace{-3ex}
   \includegraphics[width=.25\textwidth]{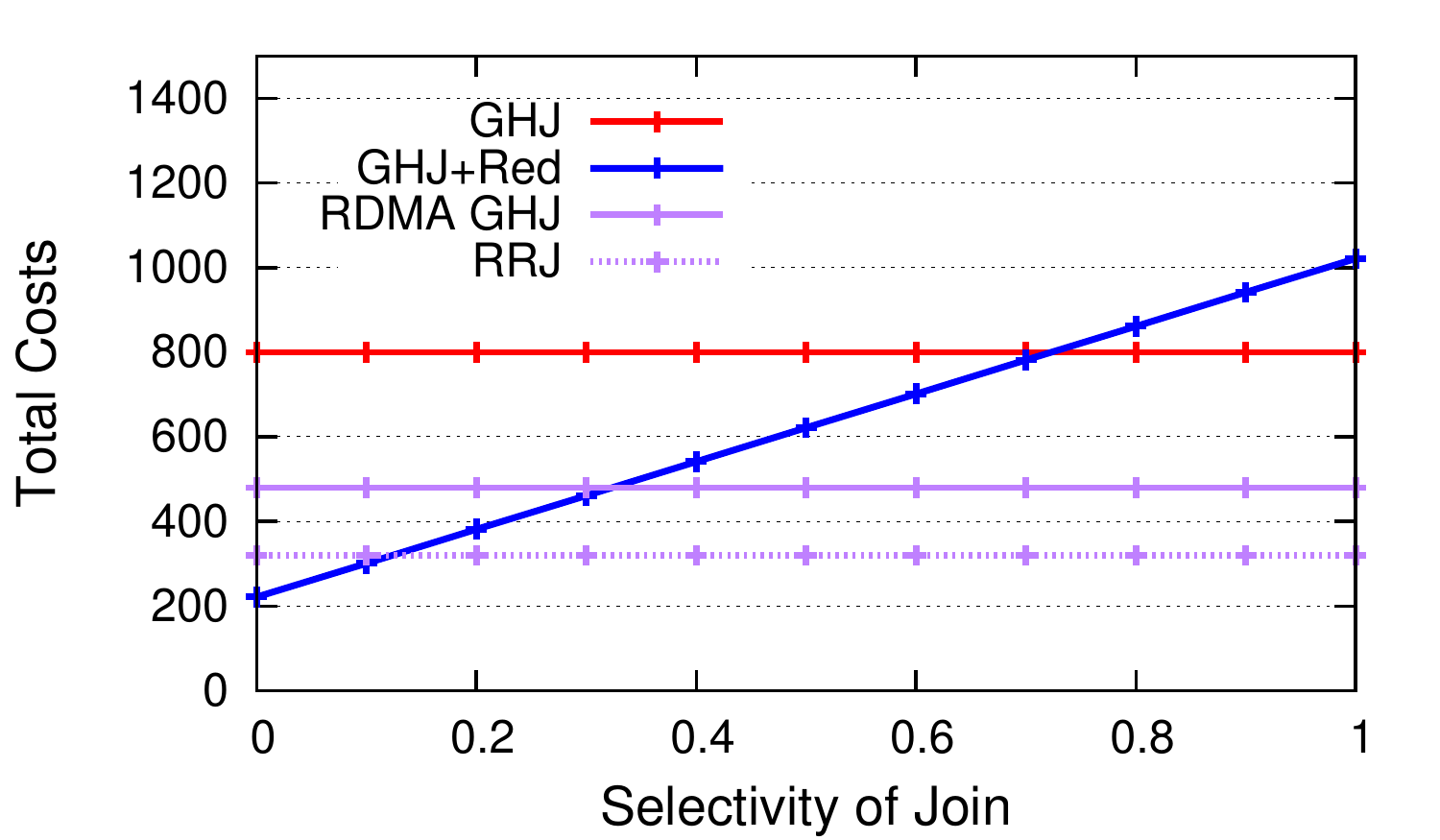}
   \label{fig:olap:cost_fast}
}
\vspace{-10pt}
\caption{Cost analysis of Joins on InfiniBand}
\vspace{-20pt}
\label{fig:olap:cost}
\end{center}
\end{figure}

Figure~\ref{fig:olap:cost} plots all the before-mentioned costs of the classical distributed joins for  different join selectivities on slow and fast networks.
For the network cost $c_{net}$ per byte, we used the idealized latency per byte from Section~\ref{sec:background} for messages of size $2$KB.
For the Bloom filters, we assume a $10\%$ error of false positives (i.e., 50\% selectivity still selects 60\% of the data).
We use $|R|=|S|=1M$ as table sizes and $w_r=w_s=8$ as tuple width.
For main memory, we assume a cost of $c_{mem} = 10^{-9}$s for accessing a single byte. 
However, the relative relationships of the different constants $c_{cpu}$, $c_{mem}$, and $c_{net}$ are more important than the absolute cost of accessing one byte from main memory.

For an IPoEth network, the results demonstrate that a semi-join reduction (GHJ+Red) almost always pays off (Figure \ref{fig:olap:cost_slow}).
However, with fast networks, the trade-offs change and thus, the optimization, for existing distributed join algorithms (Figure \ref{fig:olap:cost_fast}).
For example, already with IPoIB, the network cost is no longer the dominant cost factor. 
Only if the Bloom filter selectivity is below $sel < 0.8$ (in the graph $0.7$ because of the 10\% Bloom filter error), a semi-join reduction pays off due to reduction in join and shipping cost. 
Yet, both GHJ and GHJ+Red for IPoIB still do not take full advantage of the network capabilities of InfiniBand.
In the next section, we outline a new join algorithms  which directly take advantage of InfiniBand using RDMA.

In the following, we describe two new join algorithms that leverage the RDMA-based NAM architecture presented in Section~\ref{sec:architecture}. 
First, we redesign the GHJ to use one-sided RDMA verbs to write directly into remote memory of storage nodes for partitioning.
We call this join the RDMA GHJ.
The main goal of the partitioning phase of the RDMA GHJ for the NAM architecture is to enable data parallel execution of the join phase by the compute nodes.

The input tables for the partitioning phase are pre-fetched from the storage nodes to the compute nodes. 
Moreover, for writing the output partitions back to the storage nodes, the RDMA GHJ leverages selective signaling to overlap computation and communication. 
Thus, only the CPU of the sender is active during the partitioning phase, and the cost of partitioning reduces to $T_{part}= T_{mem}(R) + T_{mem}(S)$ because the remote data transfer for writing is executed in the background by the RNICs when using selective signaling.
Finally, the join phase also uses pre-fetching of the partitioned tables.
This leads to reduced overall join costs which renders a semi-join reduction even less beneficial when compared to the classical GHJ as shown in Figure \ref{fig:olap:cost_fast}.

While this optimization may sound trivial, however, it requires a significant redesign of the join algorithm's buffer management to work efficiently on the NAM architecture. 
Each server needs to reserve a buffer for every output partition on the storage servers to ensure that data is not overwritten during the shuffling phase.  
Moreover, the partitioning phase must be designed such that the compute nodes which execute the partitioning phase can be scaled-out independently from the storage nodes.
Describing these techniques in more detail goes beyond the scope of this paper. 

However, we can go a step further than just optimizing the partitioning phase of the GHJ to leverage RDMA.
The previously described partitioning phase of the radix join used to optimize block sizes for cache-locality is very similar to the partitioning phase of the GHJ.
Therefore, instead of trying to adjust distributed join algorithms like GHJ, we propose extending the in-memory radix join ~\cite{mmjoins:paper} to leverage RDMA directly.
We refer to this new algorithm as {\em RRJ} (RDMA Radix Join). A similar algorithm was recently presented in \cite{rdmajoin}. However, unlike our algorithm, their join has been optimized for a shared-nothing architecture while our RRJ algorithm is optimized for the NAM architecture, enabling an efficient scale-out by adding additional compute servers.

\vspace*{0pt}
\subsection{RDMA-based Join Algorithms}
\label{sec:olap:join}

\begin{figure}
\begin{center}
\subfigure[Join]{
   \hspace{-7ex}
   \includegraphics[width=.25\textwidth]{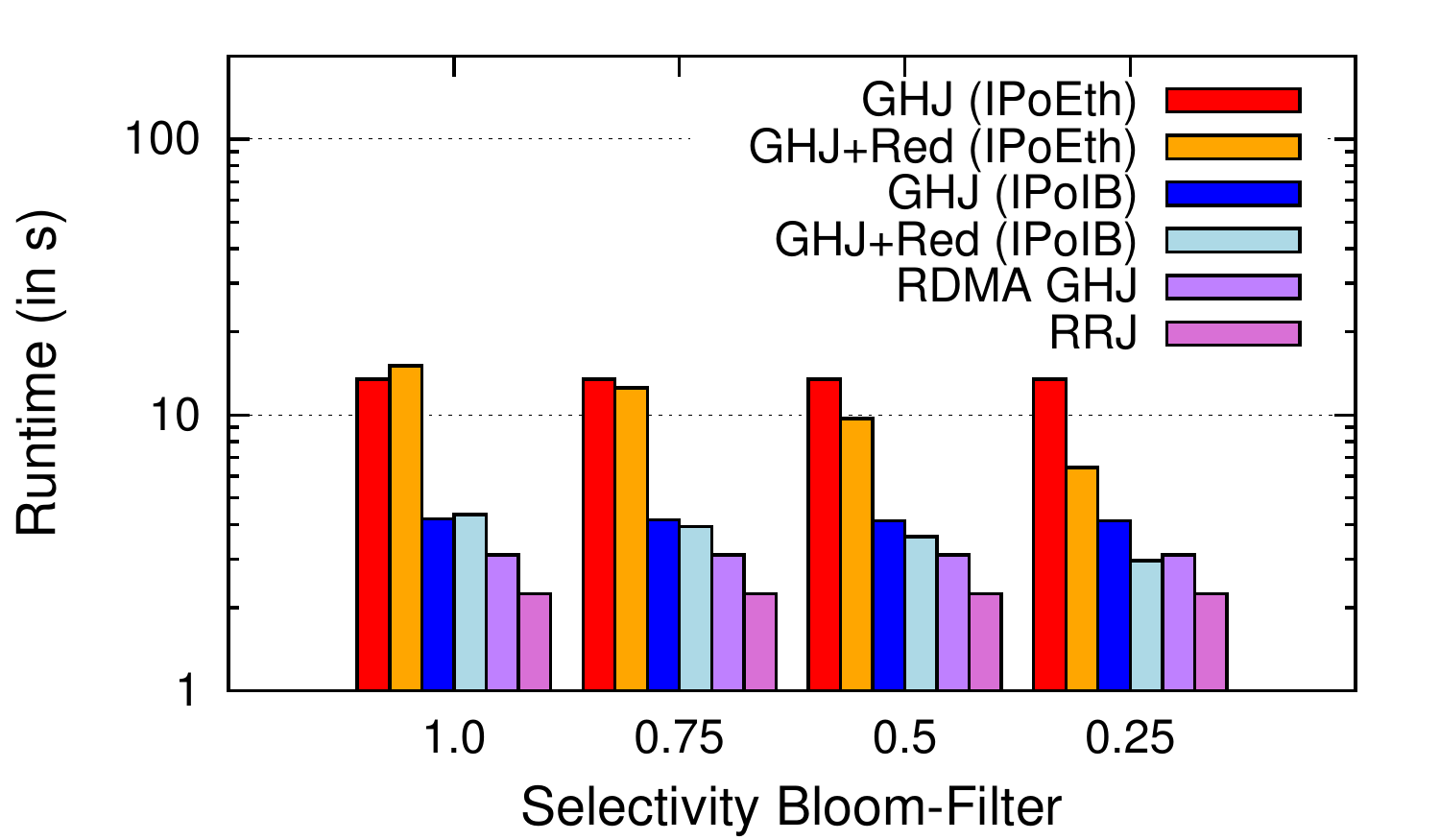}
   \label{fig:exp3join}
 }
\subfigure[Aggregation]{
   \hspace{-3ex}
   \includegraphics[width=.25\textwidth]{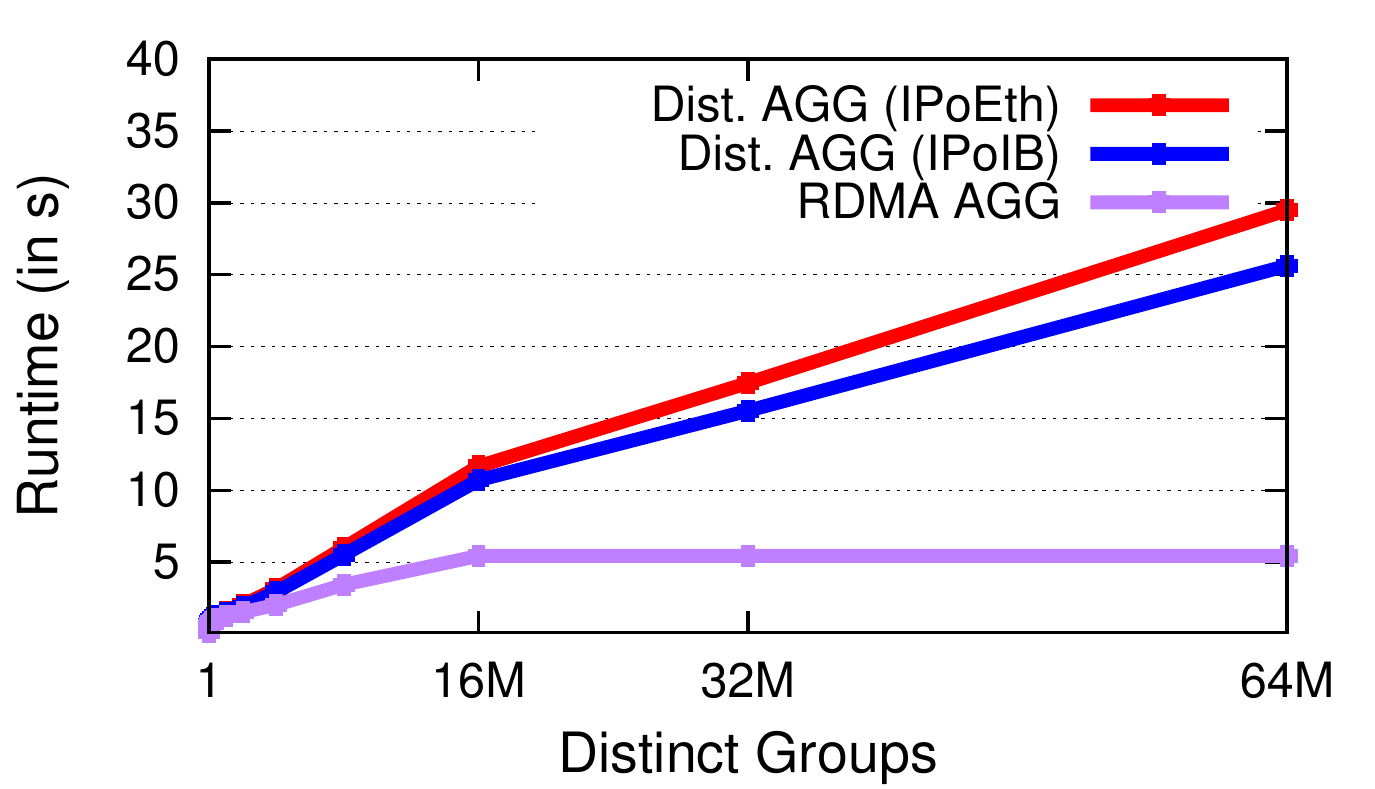}
   \label{fig:exp4agg}
}
\vspace*{-10pt}
\caption{Classical vs. RDMA-optimized}
\label{fig:exp3}
\end{center}
\vspace*{-20pt}
\end{figure}

Our new RRJ algorithm uses remote software managed buffers for the partition phase. Software managed buffers for the single-node radix join are presented in \cite{mmjoins:paper} to achieve a high fan-out of its radix-partitioning phase and avoid multiple passes. 
RRJ adopts this idea to work optimally in the NAM architecture with RDMA by applying the following changes: 
(1) buffers are copied in the background to storage nodes using selective signaled WRITEs; and (2) buffer sizes are optimized to leverage the full bandwidth of RDMA.
Our micro-benchmarks in Section~\ref{sec:background:micro} show that $2KB$ messages saturate the InfiniBand bandwidth.
Moreover, the fan-out of the remote radix-partitioning phase is selected such that all buffers fit into the L3 cache of the CPU.

Note that the resulting RRJ algorithm is not a straightforward extension of the radix join.
For example, our current implementation uses manually allocated RDMA-enabled memory in the buffer and the storage nodes. 
In a redesigned distributed DBMS, a major challenge is to manage global memory allocation efficiently without imposing a performance penalty on the critical path of a distributed join algorithm.

Assuming that the network cost is similar to the memory cost and that one partitioning pass is sufficient when using software managed buffers, the RRJ algorithm has a total expected cost of:

\vspace*{0ex}
\begin{equation*}\small
\begin{split}
T_{RRJ} &= 2 \cdot c_{mem} \cdot ( w_r \cdot |R| +  w_s \cdot |S|)
\end{split}
\label{eq:rrj:final}
\end{equation*}

The results of the cost analysis of both algorithms, the RDMA GHJ and the RRJ, is shown in Figure~\ref{fig:olap:cost_fast} and demonstrates that the popular semi-join reduction for distributed joins only pays off in corner cases (i.e., for very, very low join selectivities). 

\subsection{RDMA-based Aggregation Algorithms}
\label{sec:olap:agg}
The primary concern for distributed aggregation in a shared-nothing architecture over slow networks is to avoid network communication~\cite{Ozsu:2007:PDD:1534678}.
Traditional distributed aggregation operators therefore use a hierarchical scheme.
In a first phase all nodes individually execute an aggregation over their local data partition.
In a second phase the intermediate aggregates are then merged using a global union and a post-aggregation is executed over that union.
However, this scheme suffers from two problems:
(1) Data-skew can cause individual nodes in the first phase to take much longer than other nodes to finish.
(2) A high number of distinct group-by keys lead to high execution costs of the global union and post-aggregation.

In order to tackle these issues, we present a novel RDMA-optimized aggregation operator, which implements a distributed version of a modern in-memory aggregation operator~\cite{ibmblu13,morsel14} for our NAM architecture.
In a first phase, this operator uses cache-sized hash tables to pre-aggregate data that is local to a core (thread).
Moreover, if the hash tables are full it flushes them to overflow partitions.
In our RDMA-variant of this operator we directly copy the data in the background to remote partitions while the pre-aggregation is still active.
In a second phase, individual partitions are then post-aggregated in parallel to compute the final aggregate. 
Since this operator uses fine-grained parallelism in the first phase and there are more partitions than worker threads in the second phase, it is more robust towards data-skew and a varying number of distinct group-by keys.

\subsection{Experimental Evaluation}
\label{sec:olap:eval}

We implemented all the distributed join and aggregation variants discussed before and executed them using $4$ servers ($10$ threads per node). Each node in our experiment hosted compute and a storage node with the same configuration described in Section~\ref{sec:background:micro}.

For the join workload, we used a variant of \cite{mmjoins:paper} adopted for the distributed setting: for each node we generated a partition that has the size $|R|=|S|=128$ $Million$ and a tuple width $w_r=w_s=8B$.
We generated different data sets such that the selectivity of the Bloom filter covers $0.25$, $0.5$, $0.75$, and $1.0$ to show the effect of reduced network costs.

Figure~\ref{fig:exp3join} shows the total runtime of the GHJ and \linebreak GHJ+Red over Ethernet (IPoEth) and IP over InfiniBand (IPoIB) as well as our two RDMA variants, RDMA GHJ and RRJ, over InfiniBand (RDMA) when using $8$ threads per node. 
As shown, the new RRJ algorithm significantly outperforms the other state-of-the-art join algorithms for different semi-join selectivities. 
These results are in line with our cost analysis, though the results vary slightly as caching effects and CPU effects play a more crucial role for the RDMA variants.

In a second experiment, we analyze the performance of our RDMA Aggregation (RDMA AGG) and compare it to a classical hierarchical distributed aggregation (Dist. AGG).
For the classical aggregation, we used the algorithm as described in \cite{ibmblu13,morsel14} as local aggregation operations.
For the workload, we used one table with the size $|R|=128$ $Million$ per partition.
Each tuple of $R$ has two attributes (one group-by key and one aggregation attribute)  of $4$B each resulting in a tuple width of $w_r=8B$. 
Moreover, we generated data sets with a different number of distinct values for the group-by keys ranging from $1$ to $64M$ using a uniform distribution.

Figure~\ref{fig:exp4agg} shows the results.
For the classical hierarchical aggregation (Dist. AGG), the runtime increases with the distinct number of group-by keys due to the cost of the global union the post-aggregation (i.e., the post-aggregation has to be executed over a union which produces an output with a size of $\#nodes \cdot \#groupkeys$).
While showing a similar performance for a small number of distinct group-by keys (i.e., $0.17$ms), our RDMA Aggregation (RDMA AGG) is more robust for a high number of distinct group-by keys and shows major performance gains in that case.

Both our experiments in Figure~\ref{fig:exp3join} and Figure~\ref{fig:exp4agg} show that a redesign of DBMS operators for the NAM architecture results in major benefits not only regarding the sheer performance but also regarding other aspects such as robustness.
Different from distributed operators for the shared-nothing and shared-memory architecture, our operators are optimized for the NAM architecture, thus enabling an efficient scale-out by adding additional compute servers.
Moreover, the NAM architecture also enables more efficient schemes to handle data-skew using fine-grained parallelism and work-stealing algorithms.
All these challenges need to be addressed and analyzed in detail when redesigning distributed analytical DBMSs for the NAM architecture.


%% file: related.tex
\section{Related Work}
\label{sec:related}
A major focus in the High-Performance Computing community has been the development of techniques that take advantage of modern hardware, particularly fast networks like InfiniBand~\cite{HadoopIB,RDMAHDFS,memcachedhpc}.
While the vast majority of this work is limited to specific applications, the results and gained experiences are highly relevant for developing the next generation of DBMSs for fast networks.

In this paper, we made the case that RDMA-enabled networks should directly influence distributed DBMS architecture and algorithms.
Many projects in both academia and industry have attempted to add RDMA as an afterthought to an existing DBMS \cite{hyperrdma15,RAC}.
For example, Oracle RAC~\cite{RAC} has RDMA support, including the use of RDMA atomic primitives, but was not designed from scratch to harness the full power of the network.
However, RAC does not directly take advantage of the network for transaction processing and is essentially a workaround for a legacy system.

Recent work has investigated building RDMA-aware \linebreak DBMSs \cite{DBRAMcloud,OLAPRamCloud} on top of RDMA-enabled key/value stores~\cite{RDMA_KV,ramcloud}, but transactions and query processing are an afterthought instead of first-class design considerations. 
Other systems that separate storage from compute nodes \cite{BDS3,deuteronomy15,shared-database:sigmod2015,azure,snowflake} also treat RDMA as an afterthought.
IBM pureScale~\cite{purescale} directly leverages RDMA to provide active-active scaleout for DB2 but relies on a centralized manager to coordinate distributed transactions.
On the other hand, our NAM architecture natively incorporates RDMA primitives in order to build a shared distributed memory pool with no centralized coordinator.

The proposed ideas for RDMA build upon the huge amount of work on distributed transaction protocols  (e.g., \cite{silo,GeneralizedSI,BinnigDistSI,FederatedSI,kemme}) and distributed join processing (see \cite{DistributedQP} for an overview).
While we are not aware of any other RDMA-enabled transaction protocols, previous work has explored RDMA-aware join algorithms \cite{spinningjoin,rdmajoin}.
Unlike our approach, the work in \cite{spinningjoin} still had the assumption that networks had limited bandwidth (only 1.25GB/s) and therefore streams one relation across all the nodes (similar to a block-nested loop join).
In contrast, our RRJ join (as well as the work in \cite{rdmajoin}) is an extension of the state-of-the-art in-memory join algorithms for RDMA and is significantly more efficient than the RDMA join in \cite{spinningjoin} when comparing to the published numbers.
Moreover, unlike from \cite{rdmajoin}, our join supports efficient scale-out without re-partitioning the input tables.

%% file: concl.tex
\section{Conclusion}
\label{sec:concl}
We argued that emerging fast network technologies necessitate a fundamental rethinking of the way we build distributed DBMSs. 
Our initial experiments for OLTP and OLAP workloads already indicated the potential of fully leveraging the network. 
This is a wide open research area with many interesting research challenges to be addressed, such as the trade-off between local vs. remote processing or creating simple abstractions to hide the complexity of RDMA verbs.